\documentclass{SciPost}

\hypersetup{
    colorlinks,
    linkcolor={red!50!black},
    citecolor={blue!50!black},
    urlcolor={blue!80!black}
}
\usepackage[bitstream-charter]{mathdesign}
\DeclareSymbolFont{usualmathcal}{OMS}{cmsy}{m}{n}
\DeclareSymbolFontAlphabet{\mathcal}{usualmathcal}

\fancypagestyle{SPstyle}{
\fancyhf{}
\lhead{\colorbox{scipostblue}{\bf \color{white} ~SciPost Physics Core }}
\rhead{{\bf \color{scipostdeepblue} ~Submission }}

\fancyfoot[C]{\textbf{\thepage}}
}

\usepackage[caption=false]{subfig}
\usepackage{graphicx,epsf}
\usepackage{color,xcolor}
\usepackage{physics, outlines, bm, mathtools}
\newcommand{\opm}[1]{\bm{#1}}
\newcommand{\eye}{\pmb{\mathbb{I}}}
\newcommand{\sx}{\opm \sigma^x}
\newcommand{\sy}{\opm \sigma^y}
\newcommand{\sz}{\opm \sigma^z}
\newcommand{\bseq}{\begin{subequations}}
\newcommand{\eseq}{\end{subequations}}
\newcommand{\bsplit}{\begin{split}}
\newcommand{\esplit}{\end{split}}

\newcommand{\cov}{\text{cov}}
\newcommand{\qdot}{{\pmb{\mathring{\mathcal Q}}}}
\newcommand{\wdot}{{\pmb{\mathring{\mathcal W}}}}
\newcommand{\udot}{{\pmb{\mathring{\mathcal U}}}}

\newcommand{\hsdot}{{\opm{\dot H}_S}}
\newcommand{\rhos}{{\opm \rho_S}}
\newcommand{\rhosdot}{{\dot{\opm{\rho}}_S}}
\newcommand{\htot}{\opm{H}_\text{tot}}
\newcommand{\hs}{{\opm{H}_S}}

\begin{document}

\pagestyle{SPstyle}

\begin{center}{\Large \textbf{\color{scipostdeepblue}{
Quantum Uncertainty Relations for Thermodynamic Energy Flows\\
}}}\end{center}

\begin{center}\textbf{
Pratik Sathe\textsuperscript{1,2$\star$},
Luis Pedro Garc\'ia-Pintos\textsuperscript{1 $\dagger$} and
Francesco Caravelli\textsuperscript{1 $\circ$}
}\end{center}

\begin{center}
{\bf 1} Theoretical Division, Quantum \& Condensed Matter Physics, Los Alamos National Laboratory, USA
\\
{\bf 2} Information Science and Technology Institute, Los Alamos National Laboratory, USA
% Provide email address of corresponding author(s)
\\[\baselineskip]
$\star$ \href{mailto:sathepratik@gmail.com}{\small sathepratik@gmail.com}\,,\quad
$\dagger$ \href{mailto:lpgp@lanl.gov}{\small lpgp@lanl.gov}\, \quad 
$\circ$ \href{mailto:caravelli@lanl.gov}{\small caravelli@lanl.gov}
\end{center}

\section*{\color{scipostdeepblue}{Abstract}}
\textbf{\boldmath{%
The Heisenberg uncertainty relation, which links the uncertainties of the position and momentum of a particle, has an important footprint on the quantum behavior of a physical system. 
Analogous to this principle, we propose that thermodynamic currents associated with work, heat, and internal energy satisfy their own uncertainty relations.
To formalize this idea, we represent these currents by well-defined Hermitian operators, constructed so that their expectation values match the corresponding average currents.
Because these operators generally do not commute, the resulting quantum currents differ fundamentally from their classical counterparts.
Using the Robertson-Schr\"odinger uncertainty relation, we derive various uncertainty relations that link different thermodynamic flows.
We further illustrate this approach by applying it to quantum batteries, where we derive an energy-power uncertainty relationship and show how measurements affect the fluctuations.
}}

\vspace{\baselineskip}

\noindent\textcolor{white!90!black}{%
\fbox{\parbox{0.975\linewidth}{%
\textcolor{white!40!black}{\begin{tabular}{lr}%
  \begin{minipage}{0.6\textwidth}%
    {\small Copyright attribution to authors. \newline
    This work is a submission to SciPost Physics Core. \newline
    License information to appear upon publication. \newline
    Publication information to appear upon publication.}
  \end{minipage} & \begin{minipage}{0.4\textwidth}
    {\small Received Date \newline Accepted Date \newline Published Date}%
  \end{minipage}
\end{tabular}}
}}
}

%\linenumbers
\vspace{10pt}
\noindent\rule{\textwidth}{1pt}
\tableofcontents
\noindent\rule{\textwidth}{1pt}
\vspace{10pt}

\section{Introduction}

Quantum thermodynamics is an emerging field of physics that aims to understand the non-equilibrium behavior of small quantum systems~\cite{KosloffQuantum2013, VinjanampathyQuantum2016}. 
One of the successes of this program is the derivation and refinement of the laws of thermodynamics by formalizing notions of heat, work and entropy for out-of-equilibrium open quantum systems~\cite{AlickiIntroduction2018a, StrasbergQuantum2022}.  
For example, refinements of the second law of thermodynamics, in the form of work fluctuation equalities were first proven in the classical setting~\cite{BochkovGeneral1977,JarzynskiNonequilibrium1997,JarzynskiEquilibrium1997,CrooksEntropy1999}, and were soon followed by the derivations of their quantum mechanical counterparts~\cite{KurchanQuantum2001,TasakiJarzynski2000,CampisiQuantum2011} (see Refs.~\cite{EspositoNonequilibrium2009a,CampisiColloquium2011} for reviews).

A quantum mechanical formulation of thermodynamic quantities such as heat and work is central to quantum thermodynamics.
Since work and heat are process-dependent quantities, as opposed to being state functions~\cite{callen1991thermodynamics}, they cannot be represented as quantum mechanical observables~\cite{TalknerFluctuation2007} (see also, Refs.~\cite{RoncagliaWork2014,DeffnerQuantum2016,Perarnau-LlobetNoGo2017}).
Instead, work is usually obtained using a two-point measurement procedure which is accompanied by a loss of coherence~\cite{TasakiJarzynski2000,Mukamel2003,Monnai2005,Allahverdyan2005}.
However, their associated currents, such as the work rate (i.e., power), the heat flow, and the internal energy rate can be represented as observables with associated Hermitian operators.

While the average values of these currents equal the expectation values of the corresponding operators~\cite{Alickiquantum1979a,SpohnIrreversible1978, Koslofflinear1984, Kosloffquantum1984,VenkateshQuantum2015,LiWork2013}, particularly in the case of weak system-bath coupling, the operators themselves and their higher moments or fluctuations have not been studied to our knowledge. 
(See, however, Refs.~\cite{VenkateshQuantum2015,LiWork2013}, which study the fluctuations of work due to continuous power measurements, and Ref.~\cite{SolinasWork2013}, which examines work fluctuations using integrated power.)

\begin{figure}[t]
    \centering
    \includegraphics[width=0.9\linewidth]{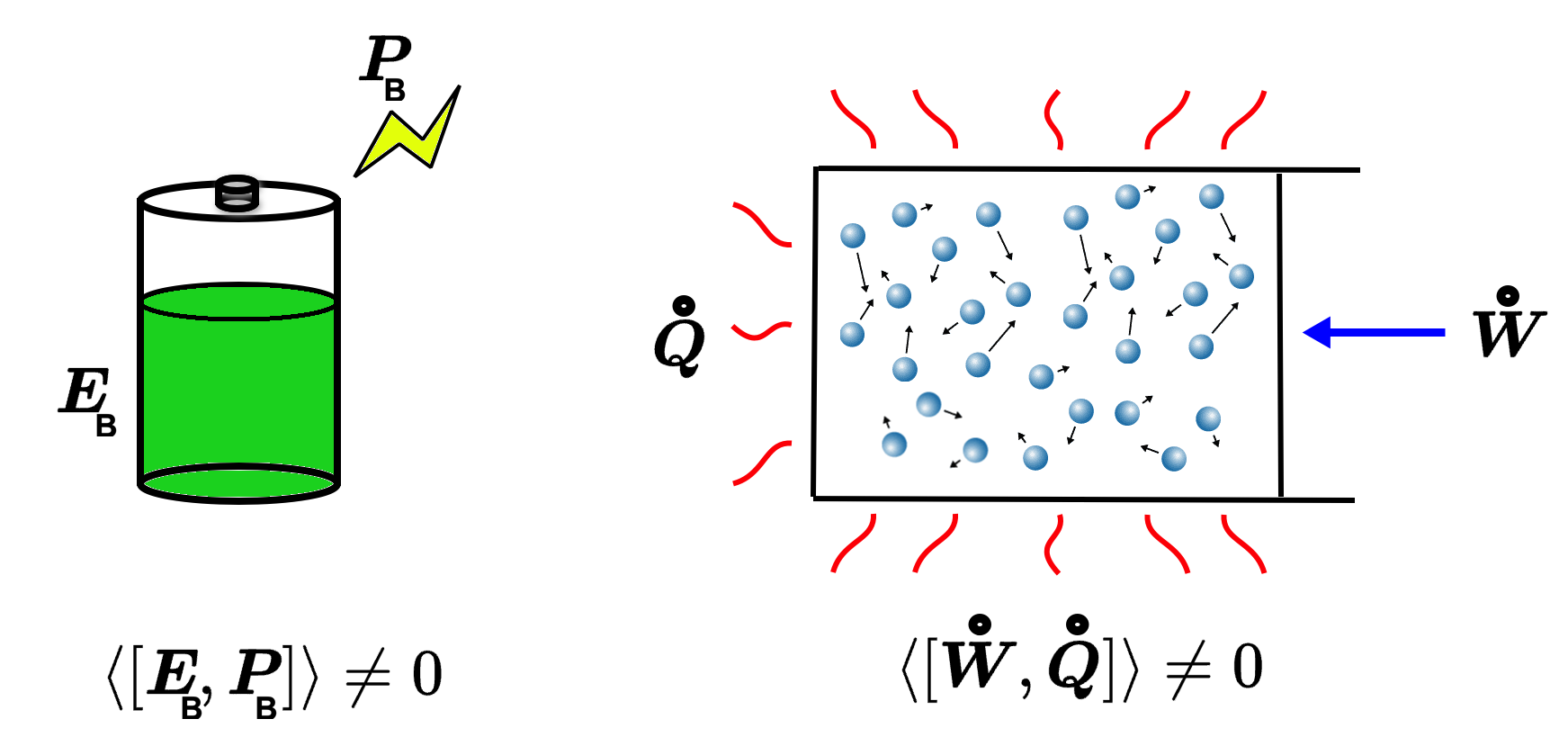}
    \caption{A pictorial representation of the non-commutativity of the energy ($\opm E_B$) and power ($\opm P_B$) operators for quantum batteries, and the power ($\wdot$) and heat flow ($\qdot$) operators in an open system. The precision of measurements of $\opm{E}_B$ and $\opm{P}_B$ for the battery (or $\wdot$ or $\qdot$ for more general open quantum dynamics) are lower bounded by the degree of non-commutativity of these operators.}
    \label{fig:fancygraph}
\end{figure}

Representing heat, work and internal energy rates as current operators yields uncertainty relations among the corresponding quantities, directly analogous to the Heisenberg position-momentum bound.
We obtain these quantum uncertainty relations (QURs) using expressions for these operators for general open quantum systems in the weak-coupling limit, as well as those whose dynamics are governed by a Markovian master equation.
Due to the generality of the Robertson-Schr\"odinger uncertainty relations these expressions are applicable at any stage in the evolution of a quantum system~\footnote{However, unlike the Heisenber uncertainty relation, the lower bound in a QUR is, in-general, state-dependent.}.
(A visual representation of the type of relationships we derive is shown in Fig.~\ref{fig:fancygraph}.)

Before proceeding, we clarify how our results relate to the thermodynamic uncertainty relations (TURs).
TURs are bounds that link the precision of thermodynamic currents to dissipation and are direct consequences of irreversibility, arising in classical~\cite{BaratoThermodynamic2015a,GingrichDissipation2016,PietzonkaFinitetime2017,HorowitzThermodynamic2020} as well as quantum thermodynamic settings~\cite{CarolloUnraveling2019,HasegawaQuantum2020,HasegawaThermodynamic2021, hasegawa2023unifying}.
In their canonical form, they assert that the thermodynamic current fluctuations are bounded from below the inverse of the total entropy production.
In contrast, we establish a \emph{complementarity principle} for various thermodynamic currents.
Achieving a small measurement variance for one current imposes a lower bound on the variance of another.
Our bounds involve products of variances and do not invoke entropy production.
They arise from operator incompatibility instead of irreversibility and therefore should not be confused with TURs.

Using various examples of open and closed quantum systems, we derive expressions for power, heat flow and energy rate operators, and numerically and analytically compare their fluctuations with the constraints implied by the operator-based uncertainty bounds.
We further highlight the utility of this approach by applying it to quantum batteries, which exploit quantum states and phenomena such as superposition and entanglement as opposed to classical batteries, which store and release energy through electro-chemical reactions.
Quantum batteries leverage the principles of quantum mechanics to create highly efficient energy storage systems (see Ref.~\cite{CampaioliColloquium2023} for a recent review), and hold the promise of a quantum advantage in terms of faster charging and work-extraction~\cite{AlickiEntanglement2013,BinderQuantacell2015,CampaioliEnhancing2017,AndolinaChargermediated2018,RossiniQuantum2020,Julia-FarreBounds2020a,GyhmQuantum2022a} compared to their classical analogues, and have also been realized experimentally~\cite{QuachSuperabsorption2022,JoshiExperimental2022}.

Quantum batteries thus serve as an interesting and novel platform for exploring quantum thermodynamic phenomena.
We establish energy-power uncertainty relationships for both open and closed quantum batteries by properly defining battery energy and battery power operators.
These relations can be interpreted to mean that the uncertainty in battery energy is inversely related to the uncertainty in battery charging power. 
We also derive the average or typical uncertainty expected in various scenarios by Haar-averaging across the state of the system. 

The rest of the paper is organized as follows:
In Sec.~\ref{sec:preliminary_discussion}, we discuss the notation, introduce average heat and work rates in quantum thermodynamics, and review the Robertson-Schr\"odinger uncertainty relation.
In Sec.~\ref{sec:Operator_Repr}, we present operator representations of heat and work rates, followed by derivations of various operator-based QURs in Sec.~\ref{sec:operator_TURs}.
In Sec.~\ref{sec:quantum_batteries}, we turn to quantum batteries and derive energy-power uncertainty relations. 
In addition, we use Weingarten calculus to derive typical uncertainty values in Sec.~\ref{sec:haar}, for the uncertainty relations derived previously in the paper.
We conclude with a summary of our results and an outlook in Sec.~\ref{sec:conclusions}.

\section{Preliminary Discussion} \label{sec:preliminary_discussion}
\subsection{Notation and Setup}\label{sec:notation_and_setup}
We will denote quantum operators by bold capital letters, while classical variables will be denoted by capital letters in standard font.
The expectation value of an observable $\opm{A}$ will be denoted by $\langle \opm{A} \rangle = \Tr(\opm{\rho} \opm{A} )$ with $\opm{\rho}$ denoting the system-environment density matrix.
We will consider the cases of general open quantum dynamics described by a system-environment Hamiltonian, open Markovian quantum systems evolving according to a Lindblad equation as well as closed but not isolated quantum systems.

We first discuss the most general case of an open quantum system, i.e., one described by a Hamiltonian that determines the evolution of the system-environment composite system:
\begin{align}
    \htot(t)=\opm{H}_S(t) \otimes \eye_E + \eye_S \otimes \opm{H}_E+\opm{V}_{SE}. \label{eq:total_Hamiltonian_notation}
\end{align} 
Here, $\opm{H}_S$ ($\opm{H}_B$) operates only on the system (environment) Hilbert space $\mathcal{H}_S$ ($\mathcal{H}_E$).
The system-environment interaction $\opm{V}_{SE}$ acts on the full Hilbert space $\mathcal H_S \otimes \mathcal{H}_E$. 
The density matrix of the composite system will be denoted by $\opm{\rho}_{SE}$, while the reduced system density matrix will be denoted by $\opm{\rho}_S=\Tr_E(\opm{\rho})$.

The average internal energy of the system is
\begin{align}
    E(t)=\Tr{\hs \opm{\rho}_{SE}}=\Tr_S\{\hs \rhos \}.\label{eq:thermodynamic_energy}
\end{align}
Taking the first derivative with respect to time, we obtain
\begin{align}
    \dot E(t)=\Tr_S\{\hsdot \rhos \}+\Tr_S\{\opm{H}_{S} \rhosdot \} \label{eq:definition_energy_rate}
\end{align}
The two terms on the right hand side of Eq.~\eqref{eq:definition_energy_rate} are usually identified as the (average) rate of change of work (i.e., power) and of heat flow respectively~\cite{StrasbergQuantum2022}:
\bseq \label{eq:work_and_heat_rate_definitions_quantum}
\begin{align}
    \dot W(t)&=\Tr_S\{\hsdot \opm{\rho}_S\}, \label{eq:work_rate_definitions_quantum}\\
    \dot Q(t)&=\Tr_S\{\opm{H}_{S} \dot{\opm{\rho}}_S\}. \label{eq:heat_rate_definition_quantum}
\end{align}
\eseq
Thus, the non-equilibrium quantum first law of thermodynamics assumes the familiar form:
\begin{align}
    \dot E(t)=\dot W(t)+\dot Q(t).
\end{align}

In the case of a quantum battery, the battery energy is defined with respect to a time-independent reference Hamiltonian, called the battery Hamiltonian, which we will denote in this paper by $\opm H_0$.
In the case of an open quantum battery, we regard the battery as being the system, with the rest being the environment. In that case, $\opm H_0$ operates only on $\mathcal H_S$.
The (average) battery energy $E_B(t)$ is then given by 
\begin{align}
    E_B(t) \coloneqq \langle \opm H_0 \rangle = \Tr_S (\opm H_0 \opm \rho_S(t)). \label{eq:battery_energy}
\end{align}
The reduced density matrix $\rho_S(t)$ evolves according to the Liouville-von Neumann equation with the full system-environment Hamiltonian, or according to a Lindblad master equation [see Eq.~\eqref{eq:Lindblad_master_equation} below] in the case of open quantum Markovian dynamics.

The instantaneous charging power of a battery is also defined with respect to $\opm H_0$, so that 
\bseq
\begin{align}
    P_B(t) &\coloneqq \dv{E_B(t)}{t} \\
    &= \Tr_S (\opm H_0 \dot{\opm \rho}_S) \label{eq:battery_average_power}
\end{align}
\eseq
We note that the battery energy $E_B(t)$ and battery charging power $P_B(t)$ are distinct from thermodynamic internal energy $E(t)$ from Eq.~\eqref{eq:thermodynamic_energy} and thermodynamic power $\dot W(t)$, since the latter are defined with respect to the full system Hamiltonian $\hs$.
We will discuss these subtleties in more detail in Sec.~\ref{sec:quantum_batteries}.

We stress here that the definitions above are valid only in the weak-coupling limit, which we will restrict our discussion to throughout this manuscript. (We refer the reader to Ref.~\cite{AlickiIntroduction2018a} for a discussion on the challenges in defining various thermodynamics quantities in the strong-coupling limit.)

\subsection{The Robertson-Schr\"odinger uncertainty relation}
To derive the QURs, we will rely on the Robertson-Schr\"odinger uncertainty relationship~\cite{RobertsonUncertainty1929, schrodinger1930heisenbergschen, Senuncertainty2014}. 
It states that, for any Hermitian operators $\opm A$ and $\opm B$,
\begin{align}
    \sigma_{\opm A}^2 \sigma_{\opm B}^2\geq \frac{1}{4} |\left\langle[\opm A,\opm{B}] \right\rangle|^2+\abs{\cov(\opm{A},\opm{B})}^2,\label{eq:robsch}
\end{align}
where
\begin{align} \cov(\bm{A},\bm{B})\coloneqq\frac{1}{2}\langle \{\bm{A},\bm{B}\}\rangle-\langle \bm{A}\rangle \langle \bm{B}\rangle 
\end{align}
is the covariance of the two operators, and $\sigma_{\opm A}^2 = \cov(\bm{A},\bm{A})$ is an operator's variance.
The square and curly brakets denote the commutator and anti-commutator of two operators respectively.
Eq.~\eqref{eq:robsch} with only the commutator term on the right hand side is known as the Robertson uncertainty relation.

The most well-known application of the Robertson relation is Heisenberg's uncertainty principle, which relates the uncertainties between the momentum and position of a particle: $\sigma_{\opm x} \sigma_{\opm p} \geq \hbar /2$.

\section{Operator Representation of Thermodynamic Flows} \label{sec:Operator_Repr}
In this section, we discuss the operator representations of thermodynamic power, heat flow, and the rate of change of internal energy of a quantum system.

\subsection{Thermodynamic Flow Operators From System-Environment Hamiltonians}\label{sec:heat_power_operators}
Consider an open quantum system described by a time-dependent Hamiltonian with a system Hamiltonian $\opm{H}_S(t)$ and an environment Hamiltonian $\opm{H}_E$. 
If their interaction is mediated by an interaction term $\opm{V}_{SE}$, the dynamics of the joint state $\opm{\rho}_{SE}(t)$ of $S$ and $E$ is determined by the total Hamiltonian given in Eq.~\eqref{eq:total_Hamiltonian_notation}.
(Often, only the system Hamiltonian $\hs$ is time-dependent, an assumption we make here. 
However, all our definitions and QUR derivations are valid even with time-dependent $\opm V_{SE}$ and $\opm H_E$.)

Eqs.~\eqref{eq:work_and_heat_rate_definitions_quantum} leads to natural definitions of the \emph{heat flow operator} $\qdot$ and the \emph{power operator} $\wdot$, defined by the property that the average heat rate and average power are equal to the expectation values of the corresponding operators.
That is, we impose that $\qdot$ and $\wdot$ satisfy
\bseq
\begin{align}
    \dot W(t) &= \langle \wdot(t) \rangle = \Tr_{S} (\wdot(t) \opm \rho_S(t)), \label{eq:defining_condition_wdot}\\
    \text{and } \dot Q(t) &= \langle \qdot(t) \rangle = \Tr_{SE} (\qdot (t) \opm \rho_S(t)). \label{eq:defining_condition_qdot}
\end{align}
\eseq
For a system described by a Hamiltonian of the form \eqref{eq:total_Hamiltonian_notation}, the operators
\bseq \label{eq:definitions_of_qdot_and_wdot}
\begin{align}    
    \wdot(t) &= \hsdot(t),     \label{eq:Wdot}\\
    \text{and }\qdot(t) &= \frac{-i}{\hbar}[\hs(t), \opm{V}_{SE}],\label{eq:definition_of_Q_dot} 
\end{align}
\eseq
satisfy Eqs.~\eqref{eq:defining_condition_wdot} and~\eqref{eq:defining_condition_qdot}.
The internal energy rate operator $\udot$ can then be defined analogously as 
\begin{align}
    \udot = \wdot + \qdot. \label{eq:udot_is_sum_of_wdot_qdot}
\end{align}

Equation~\eqref{eq:Wdot} for $\wdot(t)$ follows from comparing Eq.~\eqref{eq:work_rate_definitions_quantum} and Eq.~\eqref{eq:defining_condition_wdot}.
To derive \eqref{eq:definition_of_Q_dot} for $\qdot(t)$, we use the Liouville–von Neumann equation and Eq.~\eqref{eq:heat_rate_definition_quantum}, which leads to
\bseq \label{eq:dQ_dt_is_expect_qdot}
\begin{align}
    \dot Q (t) &= \Tr_S \left( \hs \Tr_E(\dot{\opm{\rho}}_{SE}) \right)  \\
        &= -\frac{i}{\hbar}\Tr_{SE} \left( \opm{H}_S \big[\htot, \opm{\rho}_{SE} \big] \right) \\
        &= \Tr_{SE} \Big\{\frac{-i}{\hbar}[\opm{H}_S, \htot] \opm{\rho}_{SE} \Big\} \\
        &= \Tr_{SE} \{\qdot(t) \opm{\rho}_{SE}\} = \langle \qdot(t) \rangle,
\end{align}
\eseq
thus resulting in Eq.~\eqref{eq:definition_of_Q_dot}. 

We note that while $\wdot$ and $\qdot$ correspond to work rate and heat rate respectively, they are not, in general, time derivatives of some unspecified work and heat operators.
Similarly, $\udot$ is in general not the time derivative of the internal energy operator $\opm U$.

\subsection{Thermodynamic Flow Operators From Lindblad Master Equations}
These ideas can also be applied to the case where an open quantum system's dynamics are described by a Lindblad master equation:
\bseq \label{eq:Lindblad_master_equation}
\begin{align}
    \dv{\opm \rho}{t} &= -\frac{i}{\hbar} [\opm H_S(t), \opm \rho] + \mathcal{D}_t [\opm \rho],\label{eq:Lindblad_master_equation_time_dependent}\\
    \text{with }\mathcal{D}_t[\circ ] &= \sum_k \gamma_k(t) \left( \opm{L}_k \circ \opm{L}_k^\dagger - \frac{1}{2} \{ \opm{L}_k^\dagger \opm{L}_k, \circ \} \right), \label{eq:diffusion_superoperator}
\end{align}
\eseq
where $\gamma_k(t) > 0$.
Here, the $\opm{L}_k$s denote the Lindblad operators and $\mathcal{D}_t$ denotes the (possibly-time dependent) dissipator, which includes all contributions to the non-unitary evolution of $\opm \rho$~\cite{ScopaExact2019,DannTimedependent2018}. We will restrict the discussion to the weak-coupling regime and Lindbladian evolutions where thermal states are fixed points of the dynamics, as these assumption allow deriving notions of heat and work rates that generalize those of classical stochastic thermodynamics~\cite{LandiRevModPhys.93.035008, StrasbergQuantum2022}.

Using Eq.~\eqref{eq:work_and_heat_rate_definitions_quantum} and the cyclicity of trace, we observe that
\bseq
\begin{align}
    \wdot &\equiv \hsdot, \label{eq:work_operator_Lindblad_time_independent}\\
    \text{while } \qdot &= \mathcal{D}_t^*[\opm H_S] \label{eq:heat_operator_Lindblad_time_independent}\\
    \text{where } \mathcal{D}_t^*[\circ] &\coloneqq \sum_k \gamma_k(t) \left( \opm L_k^\dagger \circ \opm L_k - \frac{1}{2}\{ \opm L_k^\dagger \opm L_k, \circ \}\right)  \label{eq:D_star_superoperator}
\end{align}
\eseq
Thus, the heat flow operator is equal to a slight modification $\mathcal D^*$ of the dissipator $\mathcal{D}$, acting upon the system Hamiltonian $\opm H_S$. 

We can use the same formalism to study other thermodynamic quantities.
For instance, the rate of change of the von Neumann entropy of the system is:
\bseq
\begin{align}
    \dot S &=-k_B \Tr (\dot{\opm{\rho}} \log \opm{\rho}), \\
    &= -k_B \Tr (\opm{\rho} \mathcal D_t^*[\log \opm{\rho}]) \\
     &= \langle \mathring {\mathcal S}_t[\opm{\rho}] \rangle \\
     \text{where } \mathring{\mathcal{S}}_t[\circ] &= -k_B \mathcal{D}_t^*[\log \circ]. \label{eq:entropy_superoperator_Lindblad_time_independent}
\end{align}
\eseq
Thus, we see that while heat and work rates can be described by operators, the entropy rate can instead be described by an entropy rate \textit{superoperator} $\mathring{\mathcal{S}}_t$.

\subsection{Examples} \label{sec:examples_operator_representation}
Next, we use the expressions for $\qdot$ and $\wdot$ to study flows in a few illustrative examples.
For all of the numerical simulations in this paper, we use the Python software package QuTip~\cite{JohanssonQuTiP2012,JohanssonQuTiP2013}.
For the two examples presented below, we discuss numerical simulations in Appendix~\ref{app:internal_energy_derivative_numerics} and show numerically that, as expected, $\dv{\langle \opm U \rangle}{t} = \langle \udot \rangle$.

\subsubsection{Two interacting spins} \label{sec:two_interacting_spins}
Let us consider two interacting spin-half particles.
In this simple model, we consider one of the particles to be the system with the other one serving as its environment:
\begin{equation} \label{eq:two_interacting_spins_example}
    \begin{split}
    \htot  &= \opm{H}_S \otimes \eye + \opm{V}_{SE} + \eye \otimes \opm H_E \\
    \text{with } \opm H_S &= f(t) \ \sx,\\
    \opm{V}_{SE} &= g\  \sz \otimes \sz, \\
    \text{and }\opm H_E &= \sx.
    \end{split}
\end{equation}
$f(t)$ denotes a time-dependent function.
Plugging into \eqref{eq:definition_of_Q_dot} and \eqref{eq:Wdot}, we obtain:
\bseq \label{eq:power_heat_rate_for_two_interacting_spins}
\begin{align}
    \qdot &= - \frac{2}{\hbar} f(t) g \opm \sigma^y \otimes \opm \sigma^z \\
    \wdot &= \dot f(t) \sx \otimes \eye, \label{eq:wdot_for_two_spins}\\
    \text{and } \udot &= \wdot + \qdot.  \label{eq:udot_for_two_spins}
\end{align}
\eseq
These operators thus assume experimentally accessible forms, that could be measured on a quantum computer.

\subsubsection{Two interacting oscillators}
\begin{figure}[t]
    \centering
    \includegraphics[width=\linewidth]{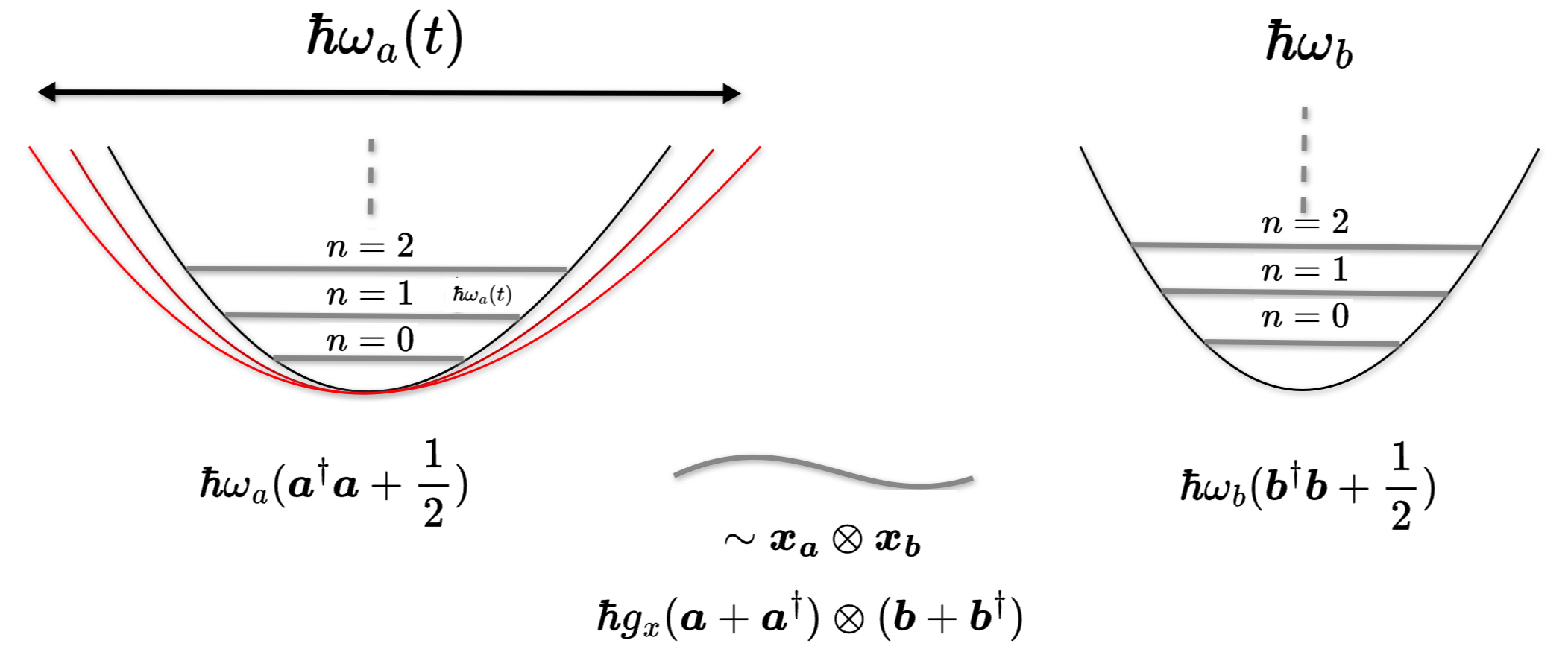}
    \caption{Pictorial representation of a system of two interacting one-dimensional harmonic oscillators [see Eq.~\eqref{eq:two_interacting_harmonic_oscillators}] . The first oscillator (left) is regarded as the system, with the second one as the environment. Work is done to the system by changing the frequency $\omega_a$ as a function of time and heat is exchanged by the system with the second oscillator due to the position-position coupling.}
    \label{fig:HarmOsci}
\end{figure}

We now consider another simple example of two one-dimensional harmonic oscillators coupled to each other (represented pictorially in Fig.~\ref{fig:HarmOsci}), with one serving as the system and the other as the environment.
Specifically, we consider~\cite{harmo0,harmo1,harmo2}:
\begin{align} 
    \bsplit
    \htot  &= \opm{H}_S \otimes \eye + \opm{V}_{SE} + \eye \otimes \opm H_E \\
    \text{with } \opm H_S &=  \frac{\opm p_a^2}{2m} + \frac{1}{2} m \omega_a(t)^2 \opm x_a^2, \\
    \opm H_E &=  \frac{\opm p_b^2}{2m} + \frac{1}{2} m \omega_b^2 \opm x_b^2,   \\
    \text{and }\opm V_{SE} &= 2g \opm  x_a \otimes \opm x_b.
    \esplit \label{eq:two_interacting_harmonic_oscillators}
\end{align}
The position (momentum) for the two oscillators are denoted by $\opm x_{a/b}$ ($\opm p_{a/b}$).

The power and heat flow operators are then easily obtained using Eq.~\eqref{eq:definitions_of_qdot_and_wdot}:
\bseq \label{eq:two_interacting_spins_wdot_qdot_xp}
\begin{align}
    \wdot &= m \omega_a \dot \omega_a \opm x_a^2 \otimes \eye _b;\\
    \qdot &= \frac{-2 g}{m} \opm p_a \otimes \opm x_b.
\end{align}
\eseq
Note that $\wdot$ is explicitly time dependent given that $\omega_a$ is time dependent. 
Furthermore, $\wdot$ is proportional to the harmonic oscillator potential energy, $\wdot  \propto \opm x_a^2$. 

\section{Quantum Uncertainty Relations for Energy Flows} \label{sec:operator_TURs}
In the previous section, we proposed definitions for heat flow and power operators for any open quantum system in the weak-coupling regime.
For some simple examples, we found that these operators assume simple, experimentally accessible forms [see Eqs.~\eqref{eq:power_heat_rate_for_two_interacting_spins} and \eqref{eq:two_interacting_spins_wdot_qdot_xp}].

An immediate consequence of the definitions of flow operators is that the fluctuations in power and fluctuations in the heat rate are related to each other by
\begin{align}
    \sigma_\wdot ^2 \sigma_\qdot ^2 \geq \frac{1}{4} \abs{\langle [\qdot, \wdot]\rangle}^2 + \abs{\cov (\qdot, \wdot)}^2\label{eq:Q_dot_W_dot_uncertainty_relation}.
\end{align}
That is, an uncertainty relation exists between the fluctuations in work and heat flows.
Eq.~\eqref{eq:Q_dot_W_dot_uncertainty_relation} follows from the Robertson-Schr\"odinger uncertainty relation applied to $\qdot$ and $\wdot$. 
(See also Ref.~\cite{Nishiyama_2024} for other uncertainty relations for Lindbladian dynamics derived from the Robertson-Schr\"odinger uncertainty relations.)

We observe from Eq.~\eqref{eq:thermodynamic_energy} that the internal energy operator $\opm U = \hs$.
Thus, $\wdot$ and $\qdot$ satisfy uncertainty relations with the internal energy and its rate of change, too, which we expand upon next.

\subsection{Internal energy uncertainty relations} \label{sec:internal_energy_uncertainty_relations}
We now derive a range of bounds on the internal energy fluctuations of open quantum systems.
Generally, $[ \opm U, \udot] \neq 0$, and consequently, using the Robertson-Schr\"odinger uncertainty relation Eq.~\eqref{eq:robsch}, we first obtain 
our first bound, which is 
\bseq
\begin{align}
    \sigma_{\opm U}^2 &\geq \frac{ \frac{1}{4}\abs{\langle [\opm U, \udot]\rangle}^2 + \abs{\cov (\opm U, \udot)}^2}{\sigma_{\udot}^2} \label{eq:sigma_u_square_intermediate}\\
    & \geq \frac{\frac{1}{4} \abs{\langle [\opm U,\udot ]\rangle}^2 + \abs{\cov (\opm U, \udot)}^2}{(\sigma_\qdot + \sigma_\wdot) ^2 - 2t_-}, \label{eq:lower_bound_on_sigma_u}
\end{align}
\eseq
Here, $t_\pm$ is defined as 
\begin{align} 
    \begin{split}
        t_\pm &= \sqrt{\frac{1}{4} \abs{\langle [\qdot, \wdot]\rangle}^2 + \abs{\cov (\qdot, \wdot)}^2} \\
        &\pm \cov (\qdot, \wdot),
    \end{split}\label{eq:t_pm}
\end{align}
To derive Eq.~\eqref{eq:lower_bound_on_sigma_u}, we use the fact that the uncertainties of $\udot$ are related to those of $\wdot$ and $\qdot$ as $\sigma_\udot^2 \leq (\sigma_\qdot+\sigma_\wdot)^2 - 2 t_-$.
(For a proof, see Appendix.~\ref{app:t_pm}.)

Similarly, an application of the Robertson-Schr\"odinger uncertainty relation to the pairs $(\opm U, \qdot)$ and $(\opm U, \wdot)$ gives us
\bseq \label{eq:TUR_U_with_qdot_wdot}
\begin{align}
    \sigma_{\opm U}^2 & \geq \frac{ \frac{1}{4}\abs{\langle [\opm U, \qdot]\rangle}^2 + \abs{\cov (\opm U, \qdot)}^2}{\sigma_{\qdot}^2},  \label{eq:lower_bound_on_sigma_u_version_qdot}\\
    \text{and }\sigma_{\opm U}^2 & \geq \frac{ \frac{1}{4}\abs{\langle [\opm U, \wdot]\rangle}^2 + \abs{\cov (\opm U, \wdot)}^2}{\sigma_{\wdot}^2}.  \label{eq:lower_bound_on_sigma_u_version_wdot}
\end{align}
\eseq

\begin{figure}[ht]
    \centering
    \includegraphics[width=0.49\linewidth]{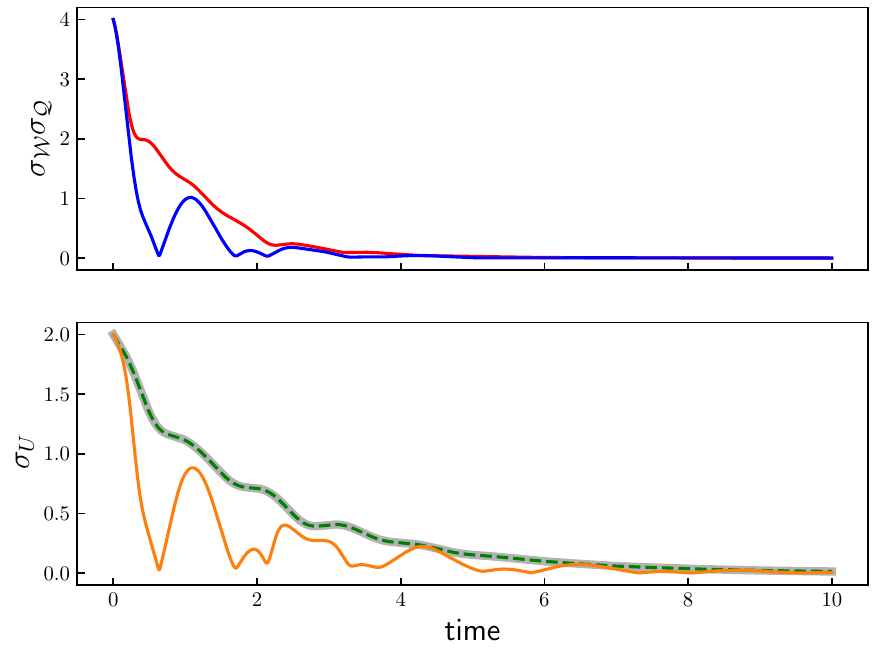}
    \includegraphics[width=0.49\linewidth]{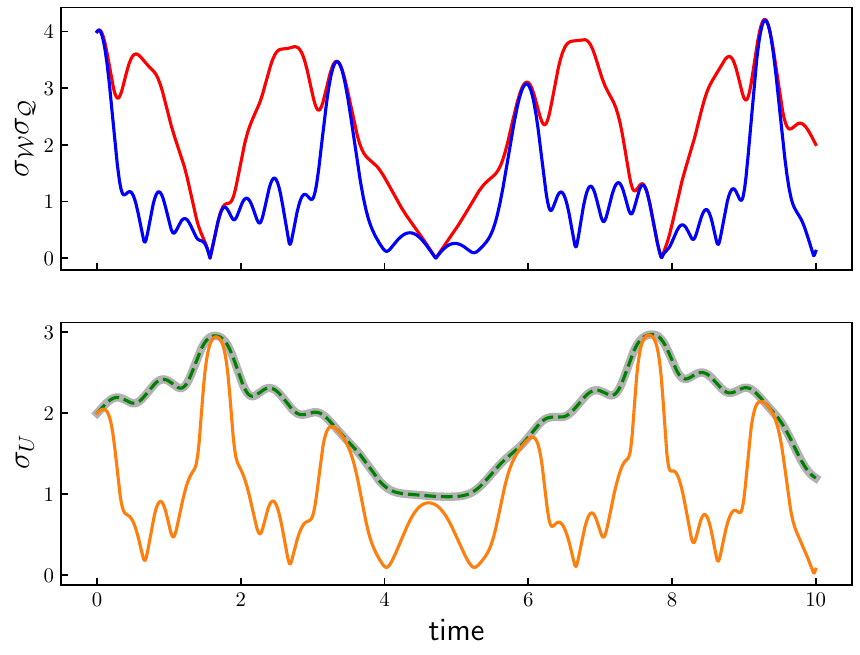}\\
    \includegraphics[width=0.75\linewidth]{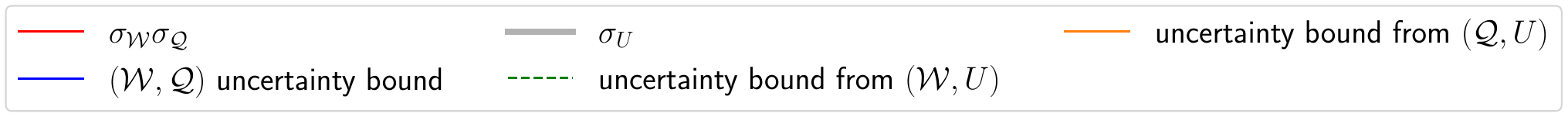}
    \caption{Numerically computed operator variances ($\sigma_\wdot, \sigma_\qdot$ and $\sigma_U$) and the corresponding lower bounds obtained using the Robertson-Schr\"odinger uncertainty relation for a system of two interacting spins with a Hamiltonian \eqref{eq:two_interacting_spins_example}. The left panel corresponds to $f(t)=2\exp(-t/2)$, while the right panel corresponds to $f(t)=\sin(t) + 2$, with $\hbar=g=1$ in both cases. The upper panel shows numerically computed $\sigma_\wdot\sigma_\qdot$ as well as the lower bound from Eq.~\eqref{eq:lower_bound_sigmas_of_qdot_wdot_two_spins}. The lower panels show $\sigma_U$ as well as lower bounds on it obtained using Eq.~\eqref{eq:lower_bound_on_sigma_u_version_qdot} and Eq.~\eqref{eq:lower_bound_on_sigma_u_version_wdot}. In all cases, the initial state was chosen to be $\ket{\psi(0)} = \ket{\uparrow} \otimes \ket{\uparrow}$, with $\ket \uparrow $ representing the up state along the $z$ direction. We note that the curve for $\sigma_{\opm U}$ coincides with the bound obtained using Eq.~\eqref{eq:lower_bound_on_sigma_u_version_wdot}.}
    \label{fig:two_interacting_spins}
\end{figure}
It is worth noting that while Ref.~\cite{DongQuantum2022} employs Robertson-Schr\"odinger uncertainty relations, the conclusions therein are substantially different and are applicable primarily to open Gaussian systems.

Typically, the commutator of two thermodynamic current operators is itself a nontrivial operator, rather than a constant as in the canonical case of position and momentum.
As a consequence, the corresponding uncertainty bound generally evolves in time.
In certain cases, however, the commutator and thus the lower bound on the product of variances, takes on a simple and physically transparent form, as illustrated below in the example of two interacting oscillators.

\subsection{Examples}
Below, we numerically verify the various bounds on $\sigma_{\opm U}$ as well as on $\sigma_\qdot \sigma_\wdot$ for the two examples discussed in the previous section.
\begin{figure}[ht]
    \centering
    \includegraphics[width=0.49\linewidth]{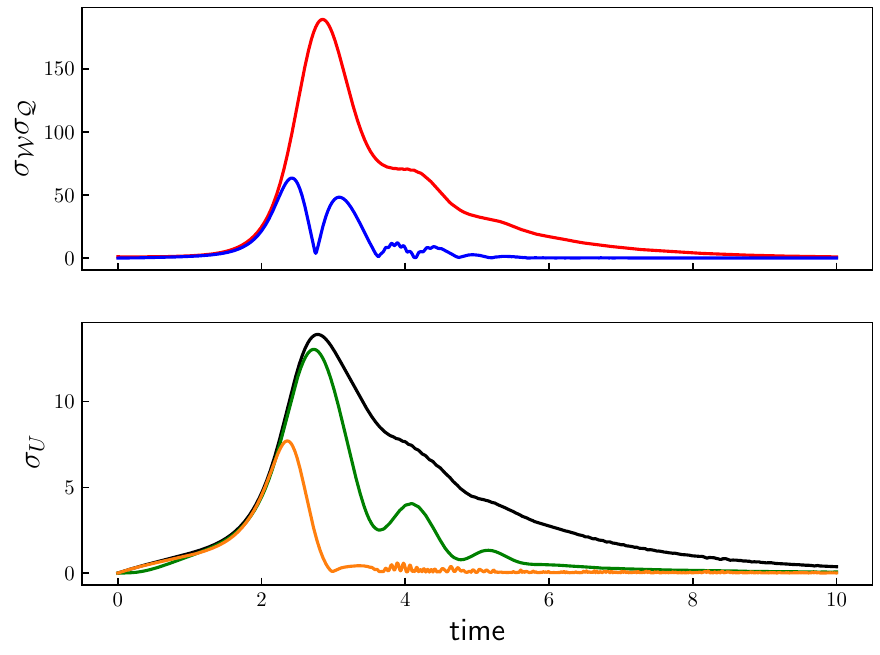}
    \includegraphics[width=0.49\linewidth]{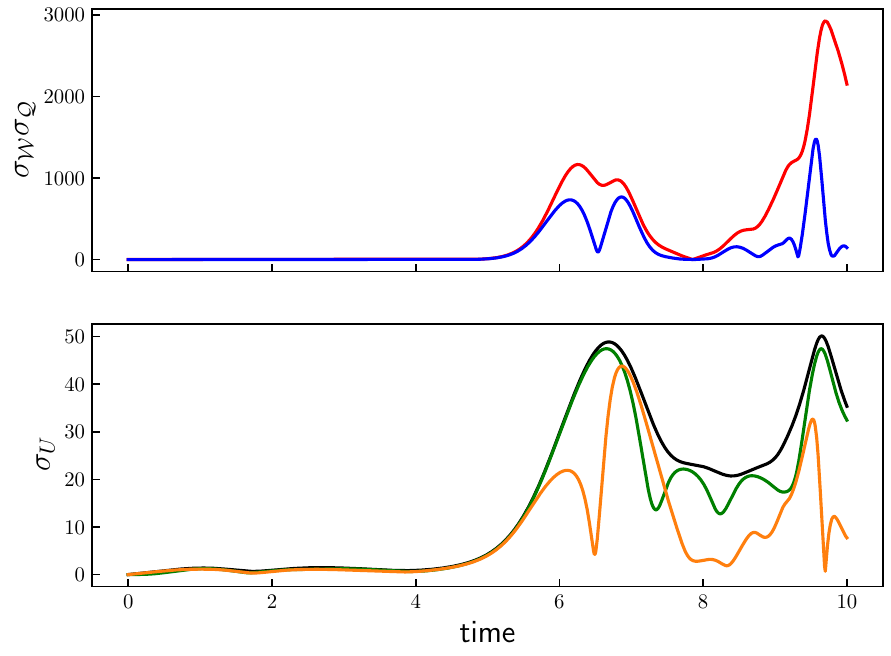}\\
    \includegraphics[width=0.75\linewidth]{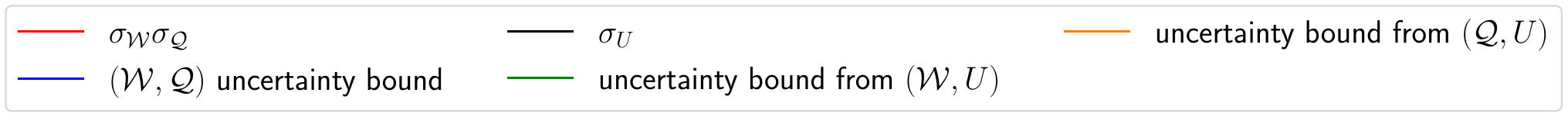}
    \caption{Numerically computed operator variances ($\sigma_\wdot, \sigma_\qdot$ and $\sigma_U$) and the corresponding lower bounds obtained using the Robertson-Schr\"odinger uncertainty relation for a system of two driven, interacting oscillators with a Hamiltonian \eqref{eq:two_interacting_harmonic_oscillators}. The left panel corresponds to $\omega_a(t)=2\exp(-t/2)$, while the right panel corresponds to $\omega_a(t)=\sin(t) + 2$, with $\hbar=g=m=\omega_b=1$ in both cases. The upper panel shows numerically computed $\sigma_\wdot\sigma_\qdot$ as well as the lower bound from Eq.~\eqref{eq:lower_bound_sigmas_of_qdot_wdot_two_oscillators}. The lower panels show $\sigma_U$ as well as lower bounds on it obtained using Eq.~\eqref{eq:lower_bound_on_sigma_u_version_qdot} and Eq.~\eqref{eq:lower_bound_on_sigma_u_version_wdot}. In all cases, the initial state was chosen to be $\ket{\psi(0)} = \ket{0} \otimes \ket{0}$, i.e. the tensor product of the ground states of the two oscillators.}
    \label{fig:two_interacting_oscillators}
\end{figure}

\subsubsection{Two interacting qubits}
Returning to the example of two interacting driven spins described by the Hamiltonian~\eqref{eq:two_interacting_spins_example}, plugging in the expressions for the power and heat rate operators \eqref{eq:power_heat_rate_for_two_interacting_spins} into \eqref{eq:Q_dot_W_dot_uncertainty_relation}, we see that the commutator and covariance terms simplify to
\begin{align}
    \abs{\frac{1}{2} \langle \{ \qdot,\wdot\}\rangle -\langle \qdot \rangle \langle \wdot \rangle }^2 &= h(t)^2   \\
    & \ \times \abs{\langle \opm \sigma^y \otimes \opm \sigma^z\rangle}^2 \abs{\langle \opm \sigma^x \otimes \eye\rangle}^2, \nonumber\\
    \text{and } \abs{\frac{1}{2i}\langle [\qdot,\wdot]\rangle}^2 &= h(t)^2 \abs{\langle {\opm V}_{SE} \rangle}^2,
\end{align}
where $h(t) = \frac{2 g f(t) \dot f(t)}{\hbar}$.  
From Eq.~\eqref{eq:Q_dot_W_dot_uncertainty_relation}, we thus have
\bseq
\begin{align}
    \sigma_{\qdot} \sigma_{\wdot} &\geq \frac{g}{\hbar} \dv{f^2(t)}{t} \sqrt {\langle \opm{ V}_{SE} \rangle^2  + \langle \sy \otimes \sz \rangle^2 \langle \sx \otimes \eye \rangle^2} \label{eq:lower_bound_sigmas_of_qdot_wdot_two_spins}\\
    &\geq \frac{g}{\hbar} \dv{f^2(t)}{t} \langle \opm{ V}_{SE} \rangle,
\end{align}
\eseq
where, the bound in the second line is simply the application of the Robertson uncertainty relation in which only the commutator term is retained.
Thus, the power-heat rate uncertainty is lower bounded by the average system-environment interaction.

Numerical simulations of this system for two choices of the driving function $f(t)$ are shown in Fig.~\ref{fig:two_interacting_spins}. We compare the values of $\sigma_\qdot \sigma_\wdot$ as well as those of $\sigma_{\opm U}$ with the corresponding lower bounds obtained from Eqs.~\eqref{eq:lower_bound_sigmas_of_qdot_wdot_two_spins}, \eqref{eq:lower_bound_on_sigma_u_version_qdot} and \eqref{eq:lower_bound_on_sigma_u_version_wdot}.

Note that for both choices of $f(t)$, $\sigma_{\opm U}$ is tightly bound by \eqref{eq:lower_bound_on_sigma_u_version_wdot}.
In this example, the bound is saturated simply because $\opm U \propto \wdot$ with a time-dependent proportionality factor, as seen by comparing Eq.~\eqref{eq:two_interacting_spins_example} and Eq.~\eqref{eq:wdot_for_two_spins}.
Consequently, in the derivation of \textbf{}the Robertson-Schr\"odinger uncertainty relation, where the Cauchy-Schwarz inequality is used, we get an equality instead of an inequality.
Furthermore, since the two operators commute, the uncertainty is fully captured by the covariance term.

The behavior in both the cases considered in Fig.~\ref{fig:two_interacting_spins}--- exponentially decaying driving (left panel) and periodic driving (right panel) can be understood qualitatively. 
In the former case, the system has no driving at all in the large $t$ limit. Consequently, $\wdot, \qdot \rightarrow 0$ as $t\rightarrow \infty$.
Naturally, the uncertainties in the work rate, heat rate as well as the internal energy (since $\opm U\rightarrow 0$ as well) all decay to zero in the long time limit.
On the other hand, with sinusoidal driving, $\wdot$, $\qdot$ and $\opm U$ all oscillate with time while $\opm \rho$ undergoes a complicated evolution. This results in an approximately oscillatory behavior of all the uncertainties.

\subsubsection{Two interacting oscillators}
For two interacting oscillators with a Hamiltonian given by Eq.~\eqref{eq:two_interacting_harmonic_oscillators}, the power-heat rate uncertainty relation Eq.~\eqref{eq:Q_dot_W_dot_uncertainty_relation} becomes
\bseq
\begin{align}
    \bsplit
    \sigma_{\qdot}^2 \sigma_\wdot^2 &\geq \hbar^2 (\omega_a \dot \omega_a)^2 \abs{\langle \opm V_{SE} \rangle}^2  + (\omega_a \dot \omega_a )^2 \label{eq:lower_bound_sigmas_of_qdot_wdot_two_oscillators} \\
    & \times \abs{2g \langle \opm x_a^2 \rangle \langle \opm p_a \opm x_b \rangle+ i\hbar \langle \opm V_{SE} \rangle - \langle \opm p_a \opm x_a \opm V_{SE}\rangle}^2
    \esplit \\
    & \geq \hbar ^2 (\omega_a \dot \omega_a)^2 \abs{\langle \opm V_{SE} \rangle}^2.
\end{align}
\eseq
Once again, the lower bound in the Robertson uncertainty relation is proportional to the rate of change of the system frequency and the average system-environment interaction.

In Fig.~\ref{fig:two_interacting_oscillators}, we numerically compute the values of $\sigma_{\wdot} \sigma_{\qdot}$ and $\sigma_{\opm U}$ and compare them against the uncertainty bounds derived above, for two different choices of $\omega_a(t)$.
Unlike the case of two interacting spins, none of the bounds is saturated identically (except at specific times), since none of the operators under consideration commute with each other.

Some interesting features can be observed in Fig.~\ref{fig:two_interacting_oscillators}.
First, the uncertainty product $\sigma_\wdot \sigma_\qdot$ starts close to zero because of the small position and momentum (which are related to the work rate and heat rate) uncertainties of the ground state of the simple harmonic oscillator.
Similarly, the internal energy uncertainty starts exactly at zero, since the initial state is an exact eigenstate of $\opm U(t=0)$.
It is also interesting, especially in the exponentially decaying driving case (left panel), that the uncertainties each initially rise to their maximum value, before again decaying to zero at infinite time. While the norms of the work rate and internal energy operators themselves monotonically decrease with time in this case, the rapid evolution of the $\opm \rho$ initially compensates, resulting the initial increase in the uncertainties. At long times, $\wdot\rightarrow 0$ implying that $\sigma_\wdot \sigma_\udot\rightarrow 0$.

%%%%%%%%%%%%%%%%%%%%%%%%%%%%%%%%%%%%%%%%%%%%%%%%%%%%%%%%%%%%%%%%%%%%%%%%%%%%%%%%%%%%%%%%%%%%%%%%%%%
\section{Quantum Batteries: Energy-Power Uncertainty Relations} \label{sec:quantum_batteries}
In this section, we discuss operatorial representations for battery energy and battery power, and derive uncertainty relations between them.

The energy stored in a quantum battery is quantified with resepct to an operator $\opm{H}_0$, also sometimes known as the battery Hamiltonian, which describes its energy levels.
The battery energy and the instantaneous battery power are measured against $\opm H_0$, as already discussed in Sec.~\ref{sec:notation_and_setup}.

\subsection{Closed quantum batteries}
Let us consider the case of a closed quantum battery evolving unitarily.
We will define an energy operator $\opm E_B$ and a power operator $\opm P_B$ whose expectation values characterize the average energy and power of the battery, respectively.
Consider a battery self-Hamiltonian $\opm H_0$ and an additional, potentially time-dependent charging potential $\opm V_S(t)$, so that the full Hamiltonian is
\begin{align}
    \htot (t) = \opm{H}_S(t) = \opm{H}_0 + \opm{V}_S(t). \label{eq:closed_quantum_battery_total_Hamiltonian}
\end{align}
The charging potential's role is to drive the battery to a state with a higher energy. 
It is assumed that $\opm V_S(t)$ is zero before and after the charging protocol. 

Since $E_B(t) = \Tr (\opm H_0 \opm{\rho} (t))$ [recall Eq.~\eqref{eq:battery_energy}], we identify $\opm H_0$ as being the battery energy operator $\opm E_B$:
\begin{align}
    \opm E_B = \opm H_0.  \label{eq:battery_energy_closed}
\end{align}
We define the battery's power operator as 
\begin{align}
    \opm P_B^c &\coloneqq -\frac{i}{\hbar}[\opm H_0,\opm V_S].\label{eq:battery_power_closed}
\end{align}
(The superscript $c$ denotes that the system is closed.)
$\opm P_B^c$ is a Hermitian operator, since both $\opm H_0$ and $\opm V_S$ are Hermitian operators.
Clearly, if $[\opm{H}_0,\opm{V}_S]=0$, $E(t)=E(0)$ for all $t$, so we $[\opm{H}_0,\opm{V}_S]\neq 0$ during the charging process of the battery.
It is straightforward to show that $P_B(t) = \langle \opm P_B^c \rangle$, with $P_B(t)$ defined in Eq.~\eqref{eq:battery_average_power}. 
The steps involved are similar to those in Eq.~\eqref{eq:dQ_dt_is_expect_qdot}.

The following uncertainty relationship follows from Eq.~\eqref{eq:robsch}:
\begin{align} \label{eq:energy_power_uncertainty_closed_battery}  
\bsplit
\sigma_{\opm E_B}^2 \sigma_{\opm P_B^c}^2 &\geq \frac{1}{4\hbar^2} \abs{ \langle [\opm{H}_0,[\opm{H}_0,\opm{V}_S]] \rangle }^2  \\
& \ + \abs{\cov (\opm H_0, \frac{-i}{\hbar}[\opm H_0, \opm V_S])}^2.
\esplit 
\end{align}
This uncertainty relationship characterizes a deviation between classical and quantum batteries. 

One may expect that the uncertainty of the energy should reduce with a reduction in the uncertainty of power, since the average value of the latter equals the derivative of the average value of the former. 
However, from Eq.~\eqref{eq:energy_power_uncertainty_closed_battery}, we conclude that this intuition is wrong, for the same reasons that it is wrong for the position and momentum of a quantum particle.
Equation~\eqref{eq:energy_power_uncertainty_closed_battery} complements other uncertainty relations that constrain quantum batteries.
References~\cite{Julia-FarreBounds2020a,Garcia-PintosFluctuations2020} prove trade-off relations between energy and extractable work fluctuations and a battery's average charging power. 

Let us note that despite superficial similarities, thermodynamic power from previous sections is different from battery power considered here.
First, we note that the battery energy operator $\opm E_B$ is different from the thermodynamic internal energy operator $\opm U$.
Unless $\opm V_S =0$, we have $\opm E_B =\opm H_0 \neq \hs = \opm U$.
Furthermore, we note that the average instantaneous battery power $P_B$ from \eqref{eq:battery_average_power} is different from the average instantaneous power $\dot W(t)$ from Eq.~\eqref{eq:work_rate_definitions_quantum}.
Since we are considering a closed quantum system, there is no heat exchange so that $\dot W(t)$ also equals the rate of change of the thermodynamic energy $P(t) \coloneqq \dot E(t)$.

While the instantaneous quantities differ, the change in the average internal energy is equal to the change in the battery energy over a charging protocol, as long as $\opm V_S(t_i) = \opm V_S(t_f) = 0$ where $t_{i}$ and $t_f$ denote the starting and ending times of the charging protocol.
Denoting the change in internal energy by $\Delta E \equiv E(t_f) - E(t_i)$ and the change in the battery energy by $\Delta E_B \equiv E_B(t_f) - E_B(t_i)$, we have
\bseq
\begin{align}
    \Delta E &= E(t_f) - E(t_i)   \\
        &= \Tr[ \hs (t_f) \opm \rho_S(t_f) ] - \Tr [ \hs (t_i) \opm \rho_S(t_i) ] \\
        &= \Tr [ \opm H_0 \opm \rho(t_f)] - \Tr[\opm H_0 \opm \rho(t_i)] \\
        &= E_B(t_f) - E_B(t_i) = \Delta E_B.
\end{align}
\eseq
Alternatively, we may write
\bseq
\begin{align}
    (\Delta E =) \int_{t_i}^{t_f} \dd t P(t) &= \int_{t_i}^{t_f} \dd t  P_B(t)\ (= \Delta E_B) \\
    \text{or }\int_{t_i}^{t_f} \dd t \langle \udot(t) \rangle &= \int_{t_i}^{t_f} \dd t \langle \opm P_B(t) \rangle 
\end{align}
\eseq
Thus, while the instantaneous rate of change of internal energy is different from the instantaneous rate of change of the battery energy, their integrals over a charging protocol are equal.

We note that the change in the battery energy is equivalent to the notion of `exclusive work'~\cite{BochkovGeneral1977} done on a system. 
In contrast, we used the notion of `inclusive work' in our definitions of thermodynamic work.
In short, exclusive work refers to the change in energy with respect to a fixed term in the system Hamiltonian, while inclusive work is computed with respect to the full system Hamiltonian.
We refer the reader to Refs.~\cite{JarzynskiComparison2007,CampisiColloquium2011} for more detailed discussions of inclusive vs exclusive work.

For completeness, we note that the thermodynamic power and heat flow operators for Eq.~\eqref{eq:closed_quantum_battery_total_Hamiltonian} are
\bseq
\begin{align}
    \wdot &= \dv{\opm H(t)}{t} = \dot{\opm V}_S(t) \\
    \qdot &= -\frac{i}{\hbar}[\opm H(t), \opm H(t)] = 0,
\end{align}
\eseq
where $\qdot=0$ since the system is closed.

\subsection{Open quantum batteries}\label{sec:openqb}
Next, we consider two descriptions of open quantum batteries. 
In one, a battery couples to an environment (which can include reservoirs and ancillae) with a total Hamiltonian:
\begin{align}
    \htot(t) = \hs \otimes \eye + \opm V_{SE} + \eye \otimes \opm H_E. \label{eq:total_Hamiltonian_notation_batt}
\end{align}
In the second case, the battery dynamics is described by a Lindblad master equation. 
In both cases, we find that the battery power operator can be split into a closed battery power operator, and one due to the interaction with the environment.

Consider the system-environment Hamiltonian in Eq.~\eqref{eq:total_Hamiltonian_notation_batt}.
The system Hamiltonian is again split into a battery Hamiltonian and a charging potential as
\begin{align}
    \opm H_S(t) = \opm H_0 + \opm V_S(t). \label{eq:open_battery_system_hamiltonian}
\end{align}
While the battery energy operator is $\opm E_B \equiv \opm H_0$, the battery power operator $\opm P_B$ is 
\bseq
\begin{align}
    \opm P_B &= \opm P_B^c \otimes \eye+ \opm P_B^o \\
    \text{with } \opm P_B^c &= -\frac{i}{\hbar} [\opm H_0, \opm V_S(t)], \\
    \text{and } \opm P_B^o &= -\frac{i}{\hbar} [\opm H_0 \otimes \eye, \opm V_{SE}].
\end{align}
\eseq
Here, the closed quantum battery operator $\opm P_B^c$ is identical to the one obtained previously [Eq.~\eqref{eq:battery_power_closed}].
The interaction with the environment contributes an additional term $\opm P_B^o$, due to the non-commutativity of the battery Hamiltonian with the system-environment interaction.

It holds that 
\begin{align}
    \dv{P_B(t)}{t} \equiv \Tr ([\opm H_0 \otimes \eye]  \dot{\opm \rho}(t)) = \langle \opm P_B \rangle.
\end{align}
As in the case of a closed quantum battery, we expect that $[\opm E_B, \opm P_B] \neq 0$, so that the energy-power uncertainty relation, in general, should include a non-zero contribution due to the commutator of the energy and power operators.

Next, consider the case of an open quantum battery described by a Lindblad master equation~\eqref{eq:Lindblad_master_equation_time_dependent}.
The system Hamiltonian $\opm H_S$ is split into a sum of a battery Hamiltonian and a charging potential, same as in  Eq.~\eqref{eq:open_battery_system_hamiltonian}.

The average battery power simplifies as follows:
\bseq
\begin{align}
    P_B(t)  &= \dv{E_B(t)}{t} = \Tr [\opm H_0 \dot{\opm \rho}(t)] \\
    &= \Tr[ \opm P_B \rho] \\
    \text{where }\opm P_B &= \opm P_B^c + \opm P_B^o, \\
    \text{with }\opm P_B^o &= \mathcal{D}_t^* [\opm H_0],
\end{align}
\eseq
with $\mathcal D^*_t$ given in Eq.~\eqref{eq:D_star_superoperator}.
 $\opm P_B^c$ is the closed quantum battery power operator that was derived in the context of closed quantum batteries, while $\opm P_B^o$ is the contribution due to dissipation.

We note that the dissipated battery power $\opm P_B^o$ is related, but in general different from the heat flow operator from Eq.~\eqref{eq:heat_operator_Lindblad_time_independent}, since
\begin{align}
    \qdot = \mathcal{D}_t^*[\opm H_S] = \opm P_B^o + \mathcal{D}_t^* [\opm V_S(t)].
\end{align}

Applying the Robertson-Schr\"odinger uncertainty relations to the $\opm E_B$ and $\opm P_B$ operators, we can thus obtain a lower bound on $\sigma_{\opm E_B} \sigma_{\opm P_B}$:
\begin{align}
    \sigma_{\opm P_B} ^2\sigma_{\opm E_B}^2 \geq \frac{1}{4}\abs{\langle [\opm P_B, \opm E_B]\rangle}^2 + \abs{\cov(\opm P_B, \opm E_B)}^2 \label{eq:batter_energy_power_uncertainty_expression}
\end{align}

\subsection{Effects of measurements and decoherence, and maximum uncertainty bounds} \label{sec:effects_measurements_decoherence}
For all the uncertainty relations derived above, the predominantly quantum contribution is due to the non-commutativity of the operators being considered.
While the covariance term also is quantum mechanical, it ceases to become so when the operators commute.
Thus, when the two operators commute, the only non-zero contribution comes from the covariance term, which simply reduces to the covariance of two random variables.

Hence, it is natural to consider the effects of decoherence on the bound, and especially on the commutator term in the uncertainty relations.
First, we note that in the Robertson-Schr\"odinger uncertainty relations for two operators $\opm A$ and $\opm B$, the commutator term involves the expectation value with respect to the state of the system.
For convenience, we define
\begin{align}
    \sqrt{\mathcal B}=\frac{1}{2} \abs{\Tr([\opm{A},\opm{B}]\opm{\rho})},
\end{align}
so that the Robertson uncertainty relation states that $\sigma_{\opm A}^2 \sigma_{\opm B}^2 \geq \mathcal{B}$.

Let us denote the spectral projectors of $\opm A$ and $\opm B$ by sets of orthogonal projectors $\{\opm{\Pi}^i_{A}\}$ and $\{ \opm{\Pi}^i_{B} \}$ respectively.
Thus, we have $[\opm{A},\opm{\Pi}^i_{A}]=0$, $[\opm{B},\opm{\Pi}^i_{B}]=0$, $\sum_i \opm{\Pi}^i_{*}=\mathbb{I}$ and $(\opm{\Pi}^i_{*})^2=\opm{\Pi}^i_{*}$.
We now define two operations, 
\bseq
\begin{align}
    \mathcal D_*(\cdot)&=\sum_i \opm{\Pi}^i_{*} \cdot \opm{\Pi}^i_{*},\\
    \text{and }\mathcal C_*(\cdot)&=(\cdot) -\sum_i \opm{\Pi}^i_{*} \cdot \opm{\Pi}^i_{*},
\end{align}
\eseq
which select the diagonal and off-diagonal elements of any operator respectively in the eigenbasis of $*$.
Then, we can write
\begin{align}
    \opm{\rho}&=\mathcal C_A(\opm{\rho})+\mathcal D_A(\opm{\rho}) =\mathcal C_B(\opm{\rho})+\mathcal D_B(\opm{\rho}).
\end{align}
Using the definitions above, it follows that
\begin{align}
     2\sqrt{\mathcal B} &=\abs{\Tr([\opm{A},\mathcal C_{A}(\opm{\rho})]\opm{B})}= \abs{\Tr([\opm{B},\mathcal C_{B}(\opm{\rho})]\opm{A})},
\end{align}
i.e., the non-commutativity of the operators $\opm{A},\opm{B}$ depends only on the off-diagonal elements of the density matrix in the basis of $\opm{A}$ and $\opm{B}$ respectively. 
For the case of battery energy-power uncertainty relations, the operator pairs of interest are the energy and power operators.
Measurements in the basis of the energy and power will immediately imply that post-measurement, the quantity $\mathcal B$ is zero. 
More generally, greater decoherence result in a smaller value of $\mathcal B$.
(The same holds true for other uncertainty relations, such as the power-heat flow operator TUR.)

Let us now study the maximum value attainable by $\sqrt{\mathcal B}$.
First, we note that
\begin{eqnarray}
     2\sqrt{\mathcal B}=|\Tr([\opm{A},\opm{\rho}]\opm{B})|=|\Tr([\opm{B},\opm{\rho}]\opm{A})|.
\end{eqnarray}
Using the Cauchy-Schwarz inequality, we can obtain two upper bounds on the commutator term, which can be combined in to the expression
\begin{align}
    \mathcal B&\leq \min \left( \frac{1}{4} \norm{ [\opm{A},\mathcal C_{A}(\opm{\rho})}^2_F \norm{ \opm{B}}^2_F, \frac{1}{4}\norm{[\opm{B},\mathcal C_{B}(\opm{\rho})}^2_F \norm{\opm{A}}^2_F \right),
\end{align}
where $\norm{\opm{C}}_F^2=\Tr(\opm{C}^\dagger \opm{C})$ is the Frobenius matrix norm of a matrix $\opm C$, which also equals the sum of squares of its eigenvalues if $\opm C$ is Hermitian. 
Since $\opm \rho^\dagger = \opm \rho$, we have $\Tr(\opm{\rho}^2)=\|\opm{\rho}\|^2_F=\mathcal P$, where $\mathcal P$ denotes the purity of $\opm \rho$.
Thus, if $\opm{\rho}$ is diagonal either in the basis of $\opm{A}$ or $\opm{B}$, the quantum uncertainty can go to zero. 

We also note that the commutator term can be written in terms of the coherence in terms of the $l_2$-induced norm~\cite{Baumgratz2014} in the basis of $\opm B$:
\begin{align}
    \mathbb{C}_B(\opm{A}) = \norm{\mathcal C_B(\opm{A}))}_F^2=\frac{1}{2}\sum_i \norm{[\opm{A},\opm{\Pi}_j^B]}_F^2.
\end{align}
Using Lemma 1 from Ref.~\cite{Caravelli2021}, we have the following bound:
\begin{align}
    \|[\opm{A},\opm{B}]\|_F^2\leq 4 \|\opm{A}\|_F^2 \mathbb{C}_A(\opm{B})
\end{align}
We will now use this bound in order to obtain an upper bound on the commutator term for various examples.

\subsubsection{Closed quantum battery} 
In the case of the energy-power uncertainty relation for a closed quantum battery, $\opm{A}=\opm{H_0}$ and $\opm{B}= \opm P_B^c = -\frac{i}{\hbar}[\opm{H}_0,\opm{V}_S]$.
We then obtain, 
\bseq
\begin{align}
    \mathcal B&\leq \frac{1}{4\hbar^2 } \norm{[\opm{H}_0,\mathcal C_{H_0}(\opm{\rho}_S)}^2_F \norm{[\opm{H}_0,\opm{V}_S}^2_F \\
    &\leq \frac{4}{\hbar ^2}\norm{\opm{H}_0}^4_F \mathbb{C}_{H_0}(\opm{\rho}_S)\mathbb{C}_{H_0}(\opm{V}_S)
\end{align}
\eseq
which shows that the uncertainty is upper bounded by both the $l_2$ induced-norm coherence of the interaction potential and the density matrix.

\subsubsection{Heat and work rates} 
Similar bounds can be obtained for the uncertainty for an open system's heat flux and work flow defined in (\ref{eq:definition_of_Q_dot}) and (\ref{eq:Wdot}) respectively.
We obtain
\begin{eqnarray}
    \mathcal B &\leq& \frac{4}{\hbar ^2}\|\opm{H}_S\|^4_F \mathbb{C}_{H_S}(\opm{\rho}_{SE})\mathbb{C}_{H_S}(\opm{V}_{SE})
\end{eqnarray}
which also shows that the uncertainty is zero if either $\opm{\rho}$ or $\opm{V}_{SE}$ are diagonal in the basis of the subsystem.

\subsection{Examples}
\subsubsection{Closed quantum system: A single qubit}\label{sec:singlequbit}
Consider a qubit with a total Hamiltonian  
\begin{align}
    \bsplit
    \htot &= \opm{H}_0 + \opm{V}_S(t) \\
    \text{with } \opm H_0 &= h_0 \opm{\sigma}^0 + h_3 \sz \\
    \text{and } \opm V_S(t) &= (v_0 \opm{\sigma}^0 + \vec v . \vec{\opm{\sigma}} ) \theta(t), 
    \esplit \label{eq:isolated_single_qubit_H}
\end{align}
where $h_0, h_3, v_1, v_2, v_3 \in \mathbb{R} $, and $\theta(t)$ denotes the Heaviside step function.
For $t\geq 0$, we have 
\begin{align*}
    \htot = \opm{H}_S &= \alpha_0 \opm{\sigma}^0 + \vec \alpha \cdot \vec{ \opm{\sigma}},
\end{align*}
where $\alpha_0 = h_0 + v_0, \alpha_1 = v_1, \alpha_2 = v_2, \alpha_3 = h_3 + v_3$, and $\vec \sigma \equiv \begin{pmatrix}
    \sx & \sz & \sz
\end{pmatrix}$.

\begin{figure}[t]
    \centering
    \includegraphics[width=0.49\linewidth]{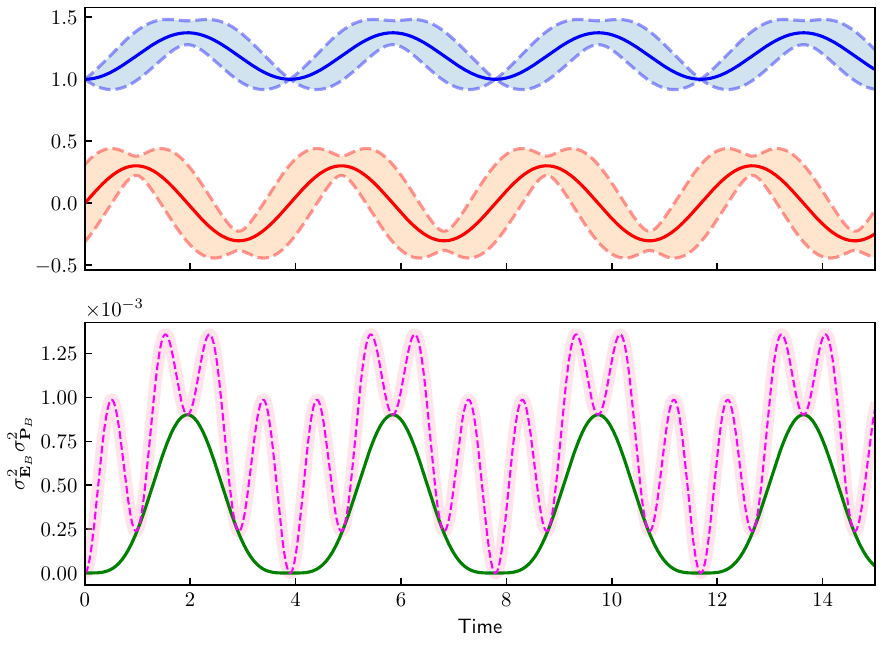}
    \includegraphics[width=0.49\linewidth]{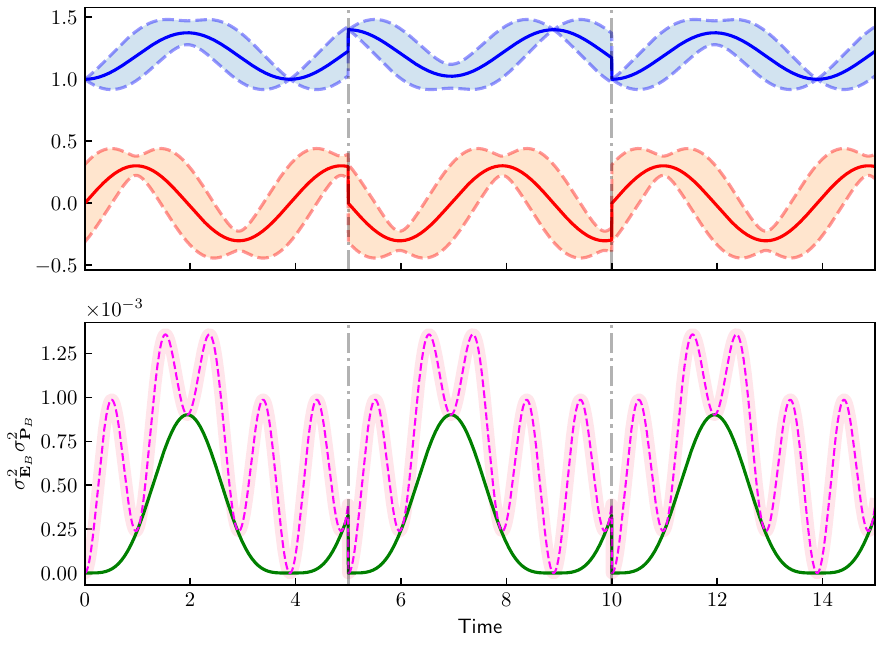}
    \\
    \includegraphics[width=0.65\linewidth]{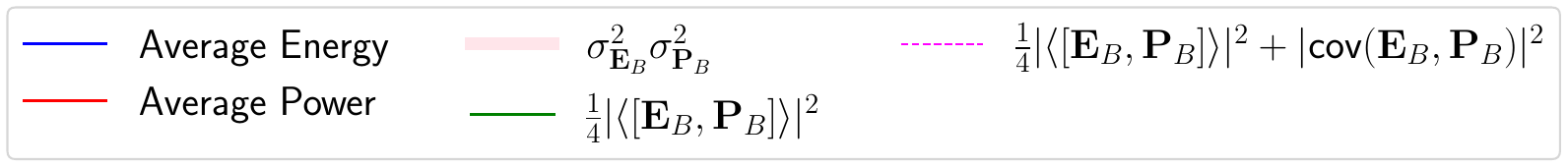}
    \caption{Energy-power uncertainty for a two-level system described by \eqref{eq:isolated_single_qubit_H} without (left figure) and with two equally-spaced measurements in the $\sigma^3$ basis (right figure). Upper panel: Evolution of average energy and average power are shown in solid lines (blue and red respectively), and the width of the shaded region represents the uncertainty $2 \sigma_{E_B}$ and $2\sigma_{P_B}$. Specifically, we plot $\langle E_B \rangle$ sandwiched between $ \langle E_B \rangle \pm \sigma_{E_B}$ and similarly for $P_B$. Lower panel: The product $\sigma_{P_B}^2 \sigma_{E_B}^2$ is shown along with the lower bound  \eqref{eq:energy_power_uncertainty_closed_battery}. We show the commutator term separately, along with the bound that includes the covariance term. The simulations correspond to the values $h_0=1.2, h_3 = 0.2$, $v_0=0$ and $\vec v = (0.5,0.6,0)$. The initial condition $\rho(t=0)$ was chosen by setting $\vec \beta = (0,0,0.5)$ in \eqref{eq:rho_0_for_isolated_spin}, i.e. the initial condition corresponds to the spin-down state in the $\sz$ basis. We set $\hbar=1$ for convenience.}
    \label{fig:epbat1q}
\end{figure} 

The time evolution operator at time $t>0$ evaluates to
\begin{align*}
    \opm{U}(0\rightarrow t) &= \mathcal{T} \exp (-i \int_0^t \opm{H}(t') \dd t') \\
    &= \exp(-i (\alpha_0 \opm{\sigma}^0 + \vec \alpha \cdot \vec{ \opm{\sigma}}) t) \\
    &= e^{-i\alpha_0 t} e^{-it(\vec{\opm{\sigma}} \cdot \vec \alpha)} \\
    &= e^{-i\alpha_0 t} \left( \opm{\sigma}^0 \cos (t\alpha) - i \frac{\sin(t\alpha)}{\alpha} (\vec{\opm{\sigma}} . \vec \alpha)\right),
\end{align*}

The density matrix $\opm \rho_0$ at $t=0$ can be written as
\begin{align}
    \opm{\rho}_0 = \frac{1}{2} \left(\opm{\sigma}^0 + \vec \beta \cdot \vec{ \opm{\sigma}} \right), \label{eq:rho_0_for_isolated_spin}
\end{align}
where $\beta \equiv \abs{\vec \beta} \leq 1$ is required for $\opm{\rho}_0$ to be a valid density matrix~\cite{nielsen2001quantum}. 
The purity is $\mathcal P=\frac{1+|\vec \beta|^2}{2}$. 

The density matrix at time $t>0$ is 
\begin{align}
    \opm{\rho}(t) = \opm{U}(0\rightarrow t) \opm{\rho}_0 \opm{U}(0\rightarrow t)^\dagger. \label{eq:exsol1q}
\end{align}
We plot the evolution along with the corresponding energy-power uncertainty relation in Fig.~\ref{fig:epbat1q}. 
In this case, the uncertainty bound is saturated at all times.
The effect of periodically measuring the energy is also shown in Fig.~\ref{fig:epbat1q}.
(An explicit expression for the commutator term within the uncertainty bound as well as comments on the consequences of energy measurement are discussed in Appendix~\ref{app:1qug}.)

The time-periodic nature of all the curves (in the absence of measurements) is natural, since the qubit's evolution is periodic, and the various battery energy and power operators are time-independent.
Similarly, in line with the discussion in Sec.~\ref{sec:effects_measurements_decoherence}, after every measurement in the basis of $\opm H_0$, $\sigma_{\opm H_0} =0$, but with $\sigma_{\opm P_B} \neq 0$. 
Since each measurement results in a complete loss of coherence, the bound also vanishes after each measurement.

\subsubsection{Spin-Boson Model} 
The spin-boson model describes a single spin interacting with an environment of harmonic oscillators, and serves as a canonical model to illustrate decoherence and dissipation in quantum systems~\cite{LeggettDynamics1987,schlosshauer2007quantum}.
We consider an open quantum battery described by this model. 

The time evolution of the spin's density matrix is determined by the Lindblad master equation:
\begin{align}
    \begin{split}
    \dv{\opm \rho_S(t)}{t} &= -\frac{i}{\hbar} [\opm H_S, \opm \rho_S(t)] \\
    & \ + \gamma \sz \opm \rho_S(t) (\sz)^\dagger - \gamma \opm \rho_S(t),
    \end{split}\label{eq:spin-boson-master-equation}
\end{align}
where $\opm H_S$ is the system Hamiltonian. We take
\bseq
\begin{align}
    \opm H_S &= {\opm H_0} + {\opm V_S(t)}. \\
    \text{with } \opm H_0 &= \alpha_3 \sz \\
    \text{and } \opm V_S(t) &= \alpha_1 \sx.
\end{align}
\eseq
The density matrix $\opm \rho_S(t)$ takes the same form as \eqref{eq:rho_0_for_isolated_spin} with a time-dependent $\vec \beta$.

Plugging into \eqref{eq:spin-boson-master-equation} yields the coupled differential equations:
\begin{align}
    \begin{split}
    \dot {\vec \beta} (t) &= \frac{1}{\hbar} \left(\vec \alpha \times \vec \beta(t)\right) - \vec \gamma \cdot \vec \beta, \\
    \text{where }\vec \gamma &\coloneqq \begin{pmatrix}
        \gamma & \gamma & 0
    \end{pmatrix} ^T.
    \end{split} \label{eq:Bloch_eq_for_spin_boson}
\end{align}
The form of these equations is the same as the Bloch equations~\cite{BlochNuclear1946} that arise in the context of nuclear magnetic resonance.
Exact solutions can be obtained (see, e.g., Refs.~\cite{TorreyTransient1949, SkinnerComprehensive2018}) for all parameter ranges.
It follows from the methods in Ref.~\cite{SkinnerComprehensive2018} that all the solutions $\beta(t)$ of Eq.~\eqref{eq:Bloch_eq_for_spin_boson} converge at $t\rightarrow \infty$ to $\beta(\infty)=0$.
To see this, note that Eq.~\eqref{eq:Bloch_eq_for_spin_boson} can also we written as
\begin{align}
    \begin{pmatrix}
        \dot \beta_1(t) \\ \dot \beta_2(t) \\ \beta_3(t)
    \end{pmatrix} + \underbrace{
    \begin{pmatrix}
        \gamma & \frac{\alpha_3}{\hbar} & 0 \\
        -\frac{\alpha_3}{\hbar} & \gamma & \frac{\alpha_1}{\hbar} \\
        0 & -\frac{\alpha_1}{\hbar} & 0 
    \end{pmatrix}
    }_{\Gamma} \begin{pmatrix}
        \beta_1(t) \\ \beta_2(t) \\ \beta_3(t)
    \end{pmatrix} = 0
\end{align}
The steady-state solution, obtained by setting $\dot{\vec \beta}=0$ is readily seen to be $\vec \beta =0$, regardless of the initial condition $\vec \beta(0)$.
Thus, as $t\rightarrow \infty$, we have $\rho_S \rightarrow \eye/2$.
The exact functional form of the decay of $\vec \beta (t)$ towards $0$, which can be under-damped, critically damped, or over-damped, depends on the parameter values $\gamma, \alpha_1$ and $\alpha_3$. (See for example, Ref.~\cite{SkinnerComprehensive2018}.)

\begin{figure*}[t]
    \centering
    \subfloat[\label{subfig:spin_boson_beta_evolution}]{%
        \includegraphics[width=0.49\linewidth]{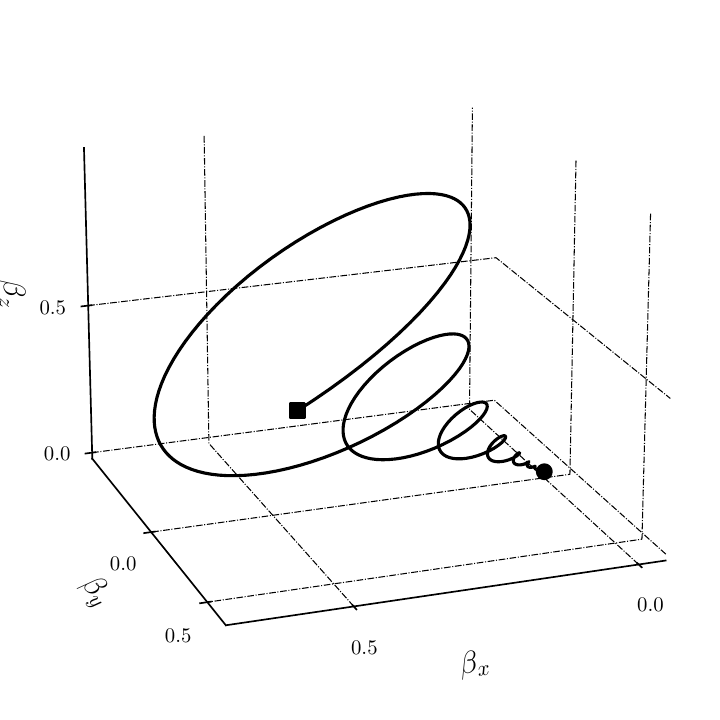}
    }
    \subfloat[\label{subfig:spin_boson_sigmas_and_bound}]{
        \includegraphics[width=0.49\linewidth]{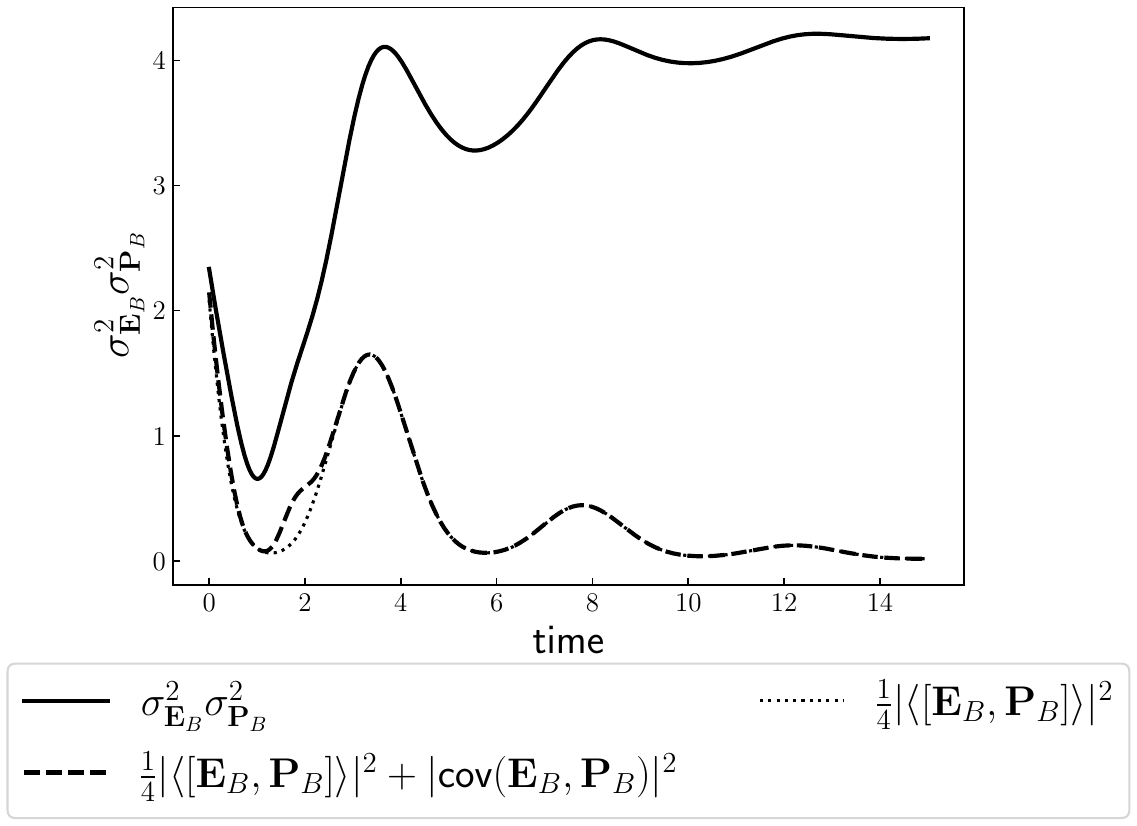}
    }
    \caption{(a) Numerically integrated solution trajectory $\beta(t)$ of the Bloch equations Eq.~\eqref{eq:Bloch_eq_for_spin_boson} for $\gamma=0.25$ and $\alpha_1=\alpha_3=\hbar=1$. The black square denotes the initial value $\vec \beta(0) = \frac{1}{\sqrt{3}}(1,1,1)^T$. The evolution (black curve) converges to the steady state, $\vec \beta(\infty)=0$ (black circle). (b) A plot of the evolution of the product $\sigma_{\opm E_B}^2 \sigma_{\opm P_B}^2$ as a function of time, along with the lower bound obtained from Eq.~\eqref{eq:batter_energy_power_uncertainty_expression}.}
    \label{fig:spin_boson_decay}
\end{figure*}

To evaluate the energy-power uncertainty relationship, let us first obtain an expression for the power operator. The contribution to the power operator due to the closed system dynamics is 
\bseq
\begin{align}
    \opm{P}_B^c&=-\frac{i}{\hbar}[\opm{H}_0,\opm{V}_S] \\
    &=-\frac{i\alpha_3\alpha_1}{\hbar}[\opm{\sigma}^z,\opm{\sigma}^x]  \\
    &=\frac{2\alpha_3\alpha_1}{\hbar}  \opm{\sigma}^y.
\end{align}
\eseq
The power contribution due to the dissipative terms is 
\bseq
\begin{align}
    \opm{P}_B^o&=\gamma \opm{L}^\dagger \opm{H}_S \opm{L}-\frac{\gamma }{2}\{\opm{H}_S,\opm{L}^\dagger \opm{L}\}\\
    &=\gamma \opm{\sigma}^z \opm{H}_S \opm{\sigma}^z-\gamma \opm{H}_S \\
    &=\gamma (-\alpha_1 \opm{\sigma}^x+\alpha_3 \opm{\sigma}^z)-\gamma (\alpha_1 \opm{\sigma}^x+\alpha_3 \opm{\sigma}^z)\\
    &=-2 \alpha_1 \gamma \opm{\sigma}^x
\end{align}
\eseq
wherein we used $(\sz)^3=\sz$ and $\sz \sx \sz=-\sz$ in the third step.
Thus, the total power operator is 
\begin{align}
    \opm P_B &= \opm P_B^c + \opm P_B^o  \nonumber \\
    &= -2\alpha_1 \gamma \sx + \frac{2\alpha_3 \alpha_1}{\hbar} \sy,
\end{align}
from which we obtain the energy-power commutator
\begin{eqnarray}
     [\opm H_0,\opm P_B]=-4i \alpha_1 \alpha_3 \left( \gamma \sy + \frac{\alpha_3}{\hbar} \sx\right).
\end{eqnarray}
We now readily obtain the various quantities in the uncertainty relationship.
First, we see that
\bseq
\begin{align}
    \sigma _{\opm E_B}^2    &= \sigma_{\opm H_0}^2 = \langle \opm H_0^2 \rangle - \langle \opm H_0 \rangle^2 \\
    &= \alpha_3^2 ( 1 - \beta_3(t)^2) \label{eq:sigma_E_square_for_spin_boson}
\end{align}
\eseq
The uncertainty in power evaluates to
\bseq
\begin{align}
    \sigma_{\opm P_B}^2 &= \langle \opm P_B^2 \rangle - \langle \opm P_B \rangle^2 \\
    &= (2\alpha_1)^2 \left[ \left(\gamma^2 - \frac{\alpha_3^2}{\hbar^2} \right) - \left( \gamma \beta_1(t) - \frac{\alpha_3\beta_2(t)}{\hbar}\right)^2 \right] \label{eq:sigma_P_square_for_spin_boson}
\end{align}
\eseq
On the other hand, we have
\bseq\label{eq:comm_and_cov_for_spin_boson}
\begin{align}
    \bsplit
    \frac{1}{4} \abs{\langle [\opm E_B, \opm P_B] \rangle}^2 &= (2\alpha_1\alpha_3)^2 \\
    & \times \left(\frac{\alpha_3 \beta_1(t)}{\hbar} + \gamma \beta_2(t)\right)^2,
    \esplit \\
    \bsplit
    \text{and } \abs{\text{cov}(\opm E_B, \opm P_B)}^2 &= (2\alpha_1\alpha_3)^2 \beta_3(t)^2   \\
    & \times \left( \frac{\alpha_3 \beta_2(t)}{\hbar} - \gamma \beta_1(t)\right)^2.
    \esplit
\end{align}
\eseq

The bound for the Robertson-Schr\"odinger uncertainty relation relation for $\opm H_0$ and $\opm P$, Eq.~\eqref{eq:batter_energy_power_uncertainty_expression} is then obtained by summing up these two terms.

While it is not the purpose of this paper to delineate the various parameter regimes, we illustrate the damping by numerically simulating the evolution for a fixed set of parameters. 
For concreteness, we choose $\gamma = 0.25$ and $\alpha_1 = \alpha_3 = \hbar$, so that the matrix $\Gamma$ becomes
\begin{align}
    \Gamma = \begin{pmatrix}
        0.25 & 1 & 0 \\
        -1 & 0.25 & 1 \\
        0 & -1 & 0 
    \end{pmatrix}.
\end{align}
The solution is then given by 
\bseq
\begin{align}
    \vec \beta(t)  &= \exp(-\Gamma t) \vec \beta(0). \\
    &= S \begin{pmatrix}
        e^{-\lambda_1 t} & 0 & 0 \\
        0 & e^{-\lambda_2 t} & 0 \\
        0 & 0 & e^{-\lambda_3 t}
    \end{pmatrix}
    S^{-1} \vec \beta(0),
\end{align}
\eseq
where $\Gamma = S \Lambda S^{-1}$ is the diagonalization of $\Gamma$, with $\lambda_i$ being the eigenvalues of $\Gamma$.
The eigenvalues are approximately $0.124$ and $0.188 \pm 1.407i$, all of which have positive real parts. 
Consequently, $e^{-\Lambda t} \rightarrow 0$ as $t\rightarrow \infty$, as expected.
We plot the evolution $\vec \beta(t)$ for an example initial value of $\vec \beta(0) = \frac{1}{\sqrt{3}}(1,1,1)^T$ in Fig.~\ref{subfig:spin_boson_beta_evolution}.

Let us now consider the various values in Eq.~\eqref{eq:batter_energy_power_uncertainty_expression} in the limit of $t\rightarrow \infty$.
Following the discussion in Sec.~\ref{sec:effects_measurements_decoherence}, expect the commutator term to approach $0$, since the qubit loses all coherence and becomes a maximally mixed state.
This is precisely what we get (along with the covariance term becoming zero) upon plugging in $\vec \beta= 0$ in \eqref{eq:comm_and_cov_for_spin_boson}. 
We also expect $\langle \opm E_B \rangle$ to equal zero since the spectrum of the battery Hamiltonian is symmetric about $0$.
$\langle \opm P_B \rangle$ also equals zero since there cannot be any charging or discharging in the steady-state limit.
However, the variances are not expected to be zero, since the second moment of both the operators is non-zero, even for the steady-state (i.e., $\vec \beta = 0$).
Indeed, for the example considered here, we get $\sigma_E=1$ and $\sigma_P = 2$ upon plugging in $\vec \beta =0$ in \eqref{eq:sigma_E_square_for_spin_boson} and \eqref{eq:sigma_P_square_for_spin_boson} respectively.
We plot the numerically obtained evolution of the uncertainty in Fig.~\ref{subfig:spin_boson_sigmas_and_bound}.

Finally, we focus on the heat flow, (thermodynamic) power and the entropy rate of Eq.~\eqref{eq:entropy_superoperator_Lindblad_time_independent}.
Since $\opm H_S$ is time-independent, the power is $\wdot = 0$.
(Consequently, the power uncertainty $\sigma_{\wdot}=0$ as well.)
On the other hand, the entropy rate and heat flow are in general non-zero, and do not commute with each other:
\begin{align}
    \bsplit
    &[\mathcal D^*(\opm{H}_S),\mathcal D^*(\log \opm{\rho})] \\  
    &=\gamma^2\Big(\opm{\sigma}^3[\opm{H}_S,\log\opm{\rho}]\opm{\sigma}^3 +[\opm{H}_S,\log\opm{\rho}] \\
    &+[\opm{H}_S\opm{\sigma}^3,(\log\opm{\rho})\opm{\sigma}^3]+[\opm{\sigma}^3\opm{H}_S,\opm{\sigma}^3(\log\opm{\rho})]\Big).
    \esplit
\end{align}
Using that $[\opm{\rho},\log \opm{\rho}]=0$, we have
\begin{align}
    \mathcal B=\frac{\gamma^4}{4} |\Tr(\opm{\sigma}^3[\opm{H}_S,\log \opm{\rho}]\opm{\sigma}^3 \opm{\rho})|^2.
\end{align}
If $\opm{V}=0$, the expression above reduces to a simpler expression
\begin{align}
    \bsplit
    &[\mathcal D^*(\opm{H}_S),\mathcal D^*(\log \opm{\rho})] \\    
    &=\gamma^2\alpha_3\Big(\opm{\sigma}^3[\opm{\sigma}^3,\log\opm{\rho}]\opm{\sigma}^3 +[\opm{\sigma}^3,\log\opm{\rho}]\Big)
    \esplit
\end{align}
and thus the commutator term in the uncertainty relation for entropy rate and heat flow is
\begin{align}
    \mathcal B=\frac{\gamma^4 \alpha_3^2}{4} |\Tr(\opm{\sigma}^3[\opm{\sigma}^3,\log \opm{\rho}]\opm{\sigma}^3 \opm{\rho})|^2
\end{align}
While the entropy rate is not a quantum mechanical observable in the sense of being a state-independent Hermitian operator, $\mathcal B$ still lower bounds the product of uncertainties of entropy rate and heat flow, since the Robertson-Schr\"odinger uncertainty relation is applicable to any pair of Hermitian operators.

\section{Typical uncertainty values}\label{sec:haar}
The power-energy uncertainty relationships in Eq.~\eqref{eq:batter_energy_power_uncertainty_expression} that constrain quantum batteries, and the power, heat flow, and internal energy uncertainty relations in Eqs.~\eqref{eq:TUR_U_with_qdot_wdot} and \eqref{eq:Q_dot_W_dot_uncertainty_relation} depend on the typically time-dependant state. 
This makes evaluating these bounds hard, since it requires full knowledge of the generally-complex time evolution of the system.
To sidestep this challenge, we evaluate the typical values of the uncertainty relations.
Specifically, we employ Weingarten calculus, which allows us to integrate over the unitary group effectively. This enables us to average out the complex behavior arising from the time evolution and focus on the statistical properties of the quantum states and operations, and look at a simpler uncertainty probe measure. Since we are interested in the quantum origin of the uncertainty, we focus only on the generalized Heisenberg uncertainty term.

First, it is easy to see that given a (scalar) random variable $a$, if $a\geq0$, then $\overline{a}\geq 0$, so that
\begin{align}
    \overline{\sigma_{\opm A}^2 \sigma_{\opm B}^2}\geq \overline{\abs{\frac{1}{2i}\langle [\opm{A},\opm{B}]\rangle}^2} \label{eq:exprv}
\end{align}
Thus, the average uncertainty $\sigma_{\opm A}^2 \sigma_{\opm B}^2$ relationship is lower bounded by the uncertainty \textit{probe}
\begin{align}
    \overline{\mathcal B}_*=\overline{\abs{\frac{1}{2i}\langle [\opm{A},\opm{B}]\rangle}^2}
\end{align}
which is the quantity we focus on in this section. 
The star at the bottom will denote the operator in which randomness is introduced. 
The average above, as we show below, is intended as the average over the unitary channel, entering either via the initial condition $\opm{\rho}_0$ or via the interactions $\opm{V}_{S}$ and $\opm{V}_{SE}$.
Depending on how the averaging is performed, we will use a different suffix of the uncertainty probe.

\subsection{Isospectral Twirling}
Before discussing the average values, we introduce the techniques we employ for the isospectral twirling of the uncertainty probe of Eq.~\eqref{eq:exprv}.
Let us briefly describe the Haar measure averages we perform in the following.
We will use the uncertainty probe as a measure of the average uncertainty given an initial condition with a given purity, in terms of the $2k$-isospectral twirling of a unitary channel is the average of $(U^{\dag}\opm{G}U)^{\otimes k}$ over $U$ sampled uniformly from the unitary group.

Let us define this average formally. Let $\mathcal{H}\simeq\mathbb{C}^d$ be a $d$-dimensional Hilbert space and let $\opm{G} \in \mathcal{H}$ be an operator. The $k$-isospectral twirling of $\opm{G}$ is defined as
\begin{align} 
\bsplit
\hat{\mathcal{R}}^{(k)}(\opm{G}) & = \int \, \dd U\, U^{\dag \otimes k}\left(\opm{G}^{k}\right)U^{\otimes k} \\
&\equiv \overline{(U^\dagger)^{\otimes {k}} \opm{G}^{\otimes k} U^{\otimes k}}^U
\esplit \label{eq:isospectral_twirling}
\end{align}
and  $\dd U$ represents the Haar measure over the unitary group $\mathcal{U}(d)$,
where the overline represents a shorthand notation for these integrals. We denote the operator being averaged without the bold notation in the following for clarity.

The quantity $\hat{\mathcal{R}}^{(k)}(\opm{G})$ is the isospectral twirling of $\opm{G}$, that is, the average over its $k$-fold channel \cite{brandao2012convergence, caravelli2020random}.
We know from Weingarten calculus that
\begin{align}
\hat{\mathcal{R}}^{(k)}(\opm{G}):=\sum_{\pi\sigma\in \mathcal S_{k}} W_g^U(\pi\sigma^{-1},d)\tr(T_{\pi}^{(k)}\opm{G}^{\otimes k})T_{\sigma}^{(k)}
\label{eq:weingartenmethod}
\end{align}
where the sum over $\pi,\sigma$ of the permutation group $\mathcal S_{k}$, and $W^U_g(\sigma^\prime,d)$ is the Weingarten coefficient associated with the permutation $\sigma^\prime$ (in the cycle representation, see \cite{caravelli2020random,Caravelli2024} for details). The key formulae used in the following are derived in Appendix \ref{app:keyfhaar}.

The average uncertainty $\sigma_{\opm A}^2 \sigma_{\opm B}^2$ relationship is lower bounded by the uncertainty \textit{probe}.
In order to obtain a `typical' value of the bound, we average over the initial condition by averaging over the random initial state, $\opm{\rho}^U_0=U^\dagger {\opm\rho}_0 U$. 
We thus have
\bseq
\begin{align}
    \overline{\mathcal B}_\rho&=\overline{\abs{\frac{1}{2i}\langle [\opm{A},\opm{B}]\rangle}^2}\equiv \overline{\abs{\frac{1}{2i}\tr{[\opm{A},\opm{B}] U^\dagger \opm{\rho}_0 U}   }^2}^U\\
    &=\frac{1}{4} \tr([\opm{A},\opm{B}]\otimes [\opm{B}^\dagger,\opm{A}] \hat {\mathcal R}^{2}(\opm{\rho}_0)),
\end{align}
\eseq
which is one of the probes we focus on in the following. 

To calculate the average uncertainty probe, we employ the technique of isospectral twirling. This involves averaging over the unitary group with respect to the Haar measure, as described previously. Specifically, for a given operator $\opm{G}$, the isospectral twirling operation $\hat{\mathcal{R}}^{(k)}(\opm{G})$ provides a way to average over all possible unitary transformations while preserving the spectrum of $\opm{G}$. By applying this operation, we can effectively average the uncertainty probe  $\overline{\mathcal B}_*$ over the unitary group, ensuring that the resulting bound is independent of the specific time evolution details of the system.

This approach simplifies the quantification of the uncertainty bounds, as it removes the need for complete knowledge of the system's time evolution. Instead, we rely on the properties of the Haar measure and the permutation operators to derive general results that hold for any unitary channel with a given spectrum. This makes the isospectral twirling technique a powerful tool for analyzing the average uncertainty in quantum systems, particularly in cases where the exact dynamics are complex or unknown.

\subsection{Closed system case}
We first assume a random initial state as a twirl of the form
\begin{eqnarray}
    \opm{\rho}_0^U=U \opm{\rho}_0 U^\dagger,
\end{eqnarray}
and because of the invariance of the measure, it can be immediately seen that time evolution must be absent in the final result. We show in Appendix~ \ref{app:closedav} that
\begin{align}
    \overline{\mathcal B}_\rho=\frac{l_s X}{(2\hbar)^2} 
\end{align}
where
\begin{align}
    X=\Tr\Big(\opm{H}_0^2\big(6 \opm{V}_S\opm{H}_0^2 \opm{V}_S-8 (\opm{H}_0 \opm{V}_S)^2+2\opm{H}_0^2\opm{V}_S^2\big) \Big)\nonumber
\end{align}
and where
\begin{align}
    l_s=\frac{d_S \mathcal P-1}{d_S(d_S^2-1)}.
\end{align}
with
\begin{align}
    \mathcal P=\Tr\opm{\rho_0}^2
\end{align}
being the initial state purity.
It is easy to see that if our initial state is fully mixed, i.e., $\mathcal P=\frac{1}{d_S}$, then $\mathcal B=0$. In the limit $ d_S\gg 1$ (and for $\mathcal P$ finite), the expression above reduces to
\begin{align}
    \overline{\mathcal B}_\rho\approx \frac{\mathcal P X}{(2 d_S\hbar)^2} 
\end{align}
For instance, for the case of the single qubit of Sec. \ref{sec:singlequbit}, with $\opm{H_0}=\alpha_3 \sz$ and $\opm{V}=v_1 \sigma^1$
\begin{align}
    X=32 \alpha_3^4v_1^2
\end{align}
and thus
\begin{align}
    \overline{\mathcal B}_\rho=\frac{4(2 \mathcal P-1)}{3 \hbar^2} \alpha_3^3 v_1^2.
\end{align}
Thus, if the initial state is completely mixed, the right-hand side of the uncertainty relationship is zero.
If we look at the exact calculation instead, it is easy to see that for $\vec \beta=0$, then $\opm{\rho}_t=\frac{\mathbb{I}}{2}$. Since $[\opm{H}_0,[\opm{H}_0,\opm{V}_S]]=4 \alpha_3^2 v_1 \opm{\sigma}^1$, we have $\langle [\opm{H}_0,[\opm{H}_0,\opm{V}_S]]\rangle_t=0$, which is consistent with the calculation above. When instead the unitary channel is applied to the interaction $\opm{V}_S$, the probe $\overline{\mathcal B}_V$ remains time-dependent, as we show in Appendix~ \ref{sec:closedtwirlint}. This makes this probe as complicated as the full dynamics to analyze. 
The results above can be connected to the $l_2$-norm coherence,
$C_{\opm A}(\opm B) := \|\,\opm B - \sum_i \opm \Pi_{\opm A}^i \opm B \opm \Pi_{\opm A}^i\,\|_F^2$, where $\opm \Pi^i_{\opm A}$ are the projectors on the $\opm A$ operator basis, connected to the off-diagonal components of an operator $\opm B$ when written in the basis of an operator $\opm A$:
\begin{eqnarray}
    D_{\opm A}(\opm B) &&:= \sum_i \opm \Pi_{\opm A}^i\, \opm B\, \opm \Pi_{\opm A}^i , 
\nonumber \\
C_{\opm A}(\opm B) &&:= \opm B - D_{\opm A}(\opm B)
        = \opm B - \sum_i \opm \Pi_{\opm A}^i\, \opm B\, \opm \Pi_{\opm A}^i .
\end{eqnarray}
(see \cite{Caravelli2021}).
Since the Robertson inequality holds pointwise, it also holds after isospectral twirling:
$\mathbb{E}_U[\sigma_{\opm A}^2\sigma_{\opm B}^2]\ge \mathbb{E}_U[\opm B^\ast]=\opm B_{\opm\rho}$.
Moreover, the commutator contribution is controlled by basis coherence:
using $\|[\opm A,\opm B]\|_F^2\le 4\|\opm A\|_F^2\,C_{\opm A}(\opm B)$, for $\opm A=\opm H_0$ and $\opm B=-\tfrac{i}{\hbar}[\opm H_0,\opm V_S]$ one finds
$B\le \tfrac{4}{\hbar^2}\|\opm H_0\|_F^4\,C_{\opm H_0}(\opm \rho_S)\,C_{\opm H_0}(\opm V_S)$,
so we obtain another way of showing that decoherence in the $\opm H_0$-eigenbasis suppresses the bound (cf.\ Ref.~\cite{Caravelli2021}).

\subsection{Open system case}
Let us now consider this probe in the case of the open system, as in Sec. \ref{sec:notation_and_setup}.
Our assumption is that the density matrix is of the from $\opm{\rho}=\opm{\rho}_S\otimes \opm{\rho}_E$,  e.g. that there is no entanglement between the subsystem and the environment, and the interaction potential is of the form $\opm{V}_{SE}=\opm{V}_S\otimes \opm{V}_E$. 
In this setup, it is easy to see that, if the unitary channel acts on the initial state of the subsystem, we have as shown in Appendix~ \ref{eq:twirlingintst} that the uncertainty probe is given by:
\begin{eqnarray}
\overline{\mathcal B}_\rho=\frac{l_s}{(2 \hbar )^2}\Tr([\opm{H}_0,[\opm{H}_0,\opm{V}_S]]^2)\cdot |\Tr\Big(\opm{\rho}_E  \opm{V}_E \Big)|^2.
\end{eqnarray}
where $l_s$ is defined analogously to before.
Thus, in this setup, we see that the probe naturally factories in the product of two terms.

If instead, the unitary channel operates on the interaction potential $U^\dagger \opm{V}_E U$. Then we have

\begin{align}
\overline{\mathcal B}_V &=\frac{\overline{\mathcal D}_E}{4}|\Tr\Big([\opm{H_0},[\opm{H_0},\opm{V}_S]]\opm{\rho}_{S}(t)\Big)|^2 
\end{align}
where, if the dimension of the bath satisfies $d_E\gg 1$
\begin{align}
    \overline{\mathcal D}_E=\frac{\Tr(\opm{V}_E^2)(1+\mathcal P_E)-d_E \Delta V^2_\infty}{d_E^2}
\end{align}
and
\begin{align}
    \Delta V^2_{\infty}=\frac{1}{d_E}\Big(\Tr(\opm{V}_E^2)-\Tr(\opm{V}_E)^2\Big).
\end{align} 
Above, $\mathcal P_E=\Tr(\opm{\rho}_E^2)$ is the purity of the bath. We see then that the probe of the uncertainty relationship is exactly the one of a closed system's quantum battery but with an overall factor that depends on the state of the bath. If the bath is in a thermal state, e.g. $\opm{\rho}_E=\frac{e^{-\beta \opm{H}_B}}{Z}$, the purity is a function of the inverse temperature $\beta$. If $\beta=0,$ $\mathcal P_E=1/d_E$, while it is $1$ at $\beta=\infty$.
When twirling only the system initial state, the typical commutator probe factorizes into a system invariant and a bath scalar,
$\opm B_{\opm\rho} \propto \mathrm{Tr}\!\big([\opm H_0,[\opm H_0,\opm V_S]]^2\big)\,|\mathrm{Tr}(\rho_E \opm V_E)|^2$,
i.e., the closed typical contribution is ``renormalized'' by the bath expectation of the coupling operator.
As in the closed case, it is controlled by subsystem-basis coherence:
$B\le \tfrac{4}{\hbar^2}\|\opm H_S\|_F^4\,C_{\opm H_S}(\opm \rho_{SE})\,C_{\opm H_S}(\opm V_{SE})$,
hence the commutator term vanishes if either the state or interaction is diagonal in the subsystem energy basis (see also \cite{Caravelli2021}).

\section{Conclusions} \label{sec:conclusions}
In this paper, we introduce quantum uncertainty relations (QURs) between pairs of thermodynamic currents for any close or open quantum system in the weak coupling regime.
To derive QURs, we define Hermitian operators $\{ \wdot, \qdot,\udot\}$  whose averages yield the work rate (i.e., power), heat flow and internal energy rate.
The existence of these operators contrasts with the difficulty in defining observables that characterize work and heat exchanges over a finite time period.
This distinction closely follows the fact that while work and heat are process-dependent, their rates are state functions.
The QURs follow from the Robertson-Schr\"odinger uncertainty relation, which relates the variances of any two quantum mechanical observables to their commutator and covariance.
These QURs thus capture fluctuations that are uniquely quantum mechanical, since non-commutativity, while inherent in quantum physics, is absent in classical physics.

For various simple models, the power and heat rate operators assume simple Hermitian forms that might be accessible experimentally.
Furthermore, for these examples, we found uncertainty bounds for power-heat flow QURs that depend on the average system-environment interaction Hamiltonian, satisfying $\sigma_{\qdot} \sigma_{\wdot} \geq a \abs{\langle \opm{V}_{SE} \rangle}$. 

To further illustrate the utility of this approach, we computed energy-power uncertainty relations for open and closed quantum batteries. 
Similar to the position-momentum uncertainty relation that underlies the Heisenberg uncertainty principle, we find the counter-intuitive result that a lower uncertainty in the battery energy is accompanied by a larger uncertainty in battery charging power, even though power is related to the time derivative of battery energy.

Next, we commented on the role of decoherence in the magnitude of the various bounds we found.
Specifically, we showed that decoherence in the eigenbasis of either of the two operators in our operator-based TURs results in the bound becoming smaller, since the expectation value of the commutator of the two operators decreases with larger decoherence, becoming zero when the state of the system becomes fully decoherent.
In order to obtain an estimate for the typical values of the uncertainty bounds, we used Weingarten calculus to derive Haar-averaged bounds.
Since our bounds require knowledge of the state of the system (i.e., the density matrix), such typical values can serve as useful guesses when the form of the density matrix cannot be solved exactly.

The most suggestive future research direction involves bridging the gap between QURs and thermodynamic uncertainty relations (TURs) by understanding scenarios where one reduces to the other.
It would be interesting to explore higher moments of these quantities and derive various fluctuation relations from an operatorial perspective for current operators.  
Another interesting direction would be to extend the definitions of power and heat flow to systems with infinite-dimensional Hilbert spaces, since the cyclic property of the trace, which we used extensively in this work, may not always be applicable to such systems.
Given the rapid development of quantum computing devices, it would be interesting to verify and probe these QURs experimentally on gate-based quantum computers and quantum annealing devices.

\section*{Acknowledgements}
We thank Tanmoy Biswas for helpful discussions.
The authors benefited from the stimulating atmosphere of the Quantum to Cosmos workshop, organized at LANL.

\paragraph{Funding information}
P.S. and F.C. acknowledge the support of NNSA for the U.S. DoE at LANL under Contract No. DE-AC52-06NA25396, and Laboratory Directed Research and Development (LDRD) for support through 20240032DR.
L.P.G.P. acknowledges support by the LDRD program of LANL under project number 20230049DR, and Beyond Moore’s Law project of the Advanced Simulation and Computing Program at LANL managed by Triad National Security, LLC, for the National Nuclear Security Administration of the U.S. DOE under contract 89233218CNA000001.

\begin{appendix}
\numberwithin{equation}{section}

\section{Numerical Simulations for examples of interacting spins and oscillators} \label{app:internal_energy_derivative_numerics}
In this appendix, we present details of numerical simulations of the two examples (interacting spins and interacting oscillators) discussed in Sec.~\ref{sec:examples_operator_representation}.

Additionally, for each example, we numerically show below that the operator corresponding to the rate of change of internal energy has an expectation value equal to the time derivative of the internal energy expectation value.
Specifically, Eq.~\eqref{eq:definition_energy_rate} in conjunction with Eq.~\eqref{eq:definitions_of_qdot_and_wdot} implies that
\begin{align}
    \dv{\langle \opm U \rangle}{t} = \langle \udot \rangle, \label{eq:udot_and_dot_u}
\end{align}
with $\opm U = \opm H_S$ denoting the internal energy of the system.
We now discuss the numerical simulations in some detail.

\subsection{Two interacting spins}
\begin{figure*}[t]
    \centering
    \includegraphics[width=0.25\linewidth]{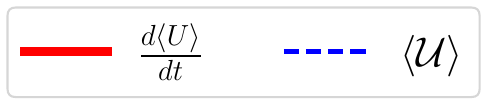}\\
    \subfloat[\label{subfig:udot_vs_dot_u_spins_sine}]{%
        \includegraphics[width=0.45\linewidth]{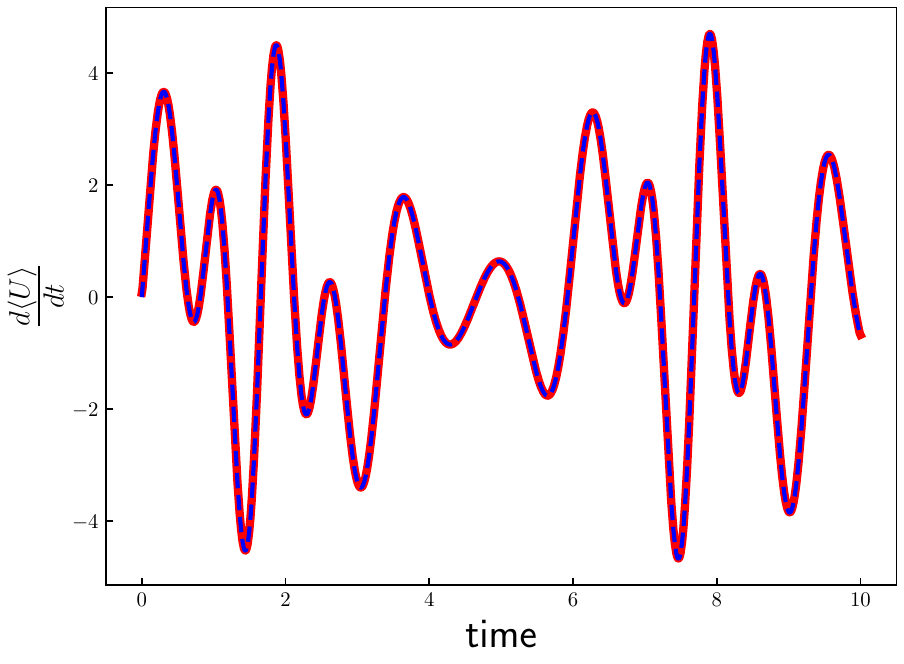}
    }
    \subfloat[\label{subfig:udot_vs_dot_u_osci_sine}]{
        \includegraphics[width=0.45\linewidth]{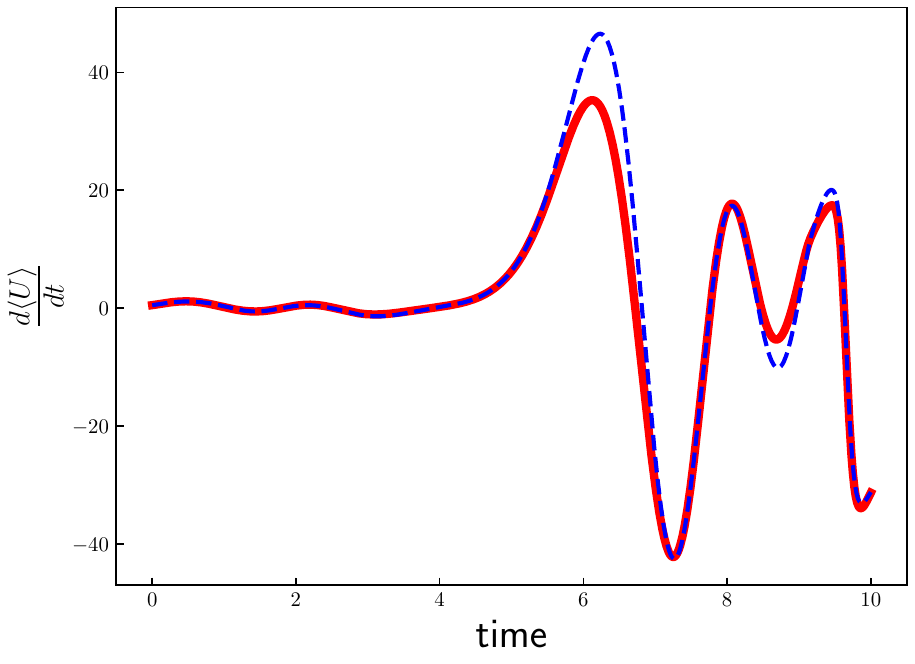}
    }
    \caption{Numerical verification of Eq.~\eqref{eq:udot_and_dot_u} for (a) two interacting spins [Eq.~\eqref{eq:two_interacting_spins_example}] and (b) two interacting harmonic oscillators [Eq.~\eqref{eq:two_interacting_harmonic_oscillators}]. For (a), we set $g=\hbar=1$ and $f(t)=\sin t+ 2$, and use the initial condition $\ket{\psi(0)} = \ket{\uparrow} \otimes \ket{\uparrow}$, with $\ket{\uparrow}$ denoting the eigenvalue $1$ eigenvector of $\sz$. For (b) we set $\omega_a(t)=\sin(t) + 2$, with $\hbar=g=m=\omega_b=1$. The initial condition at $t=0$ is chosen to be $\ket{\psi(0)}= \ket{0} \otimes \ket 0$, i.e. the tensor product of the ground states of the two oscillators. The discrepancy between the two curves in (b) arises due to numerical errors associated with truncation of the Hilbert space to low-energy subspaces of the two oscillators.}
    \label{fig:udot_vs_dot_u}
\end{figure*}

Let us revisit the example of two interacting spins with a driving term applied to one of the spins, with a Hamiltonian given by Eq.~\eqref{eq:two_interacting_spins_example}.
We find numerically that Eq.~\eqref{eq:udot_and_dot_u} is satisfied for this example, as seen in Fig.~\ref{subfig:udot_vs_dot_u_spins_sine}. 
Using numerical ODE integration, we first obtain the solution of the Schrodinger equation over a grid of 1000 equally spaced time instances between $t=0$ and $t=10$.
Next, we compute $\langle \opm U \rangle$ and its derivative $\dv{\langle \opm U \rangle}{t}$ using a standard second-order approximation, at each point of the time grid.
On the other hand, $\langle \udot \rangle$ was computed on the same time grid by computing the expectation value of $\udot$ from Eq.~\eqref{eq:udot_for_two_spins}.

As seen in {Fig.~\ref{subfig:udot_vs_dot_u_spins_sine}}, there is a close match between the time derivative of average internal energy, and the expectation value of the energy rate operator.
We note that a similar direct verification of the validity of our expressions for $\wdot$ and $\qdot$ cannot be done, since as discussed in the introduction, the work done and heat exchanged over a finite period of time are not state functions, and consequently, they are not described by quantum mechanical observables.

\subsection{Two interacting oscillators}
The Hamiltonian Eq.~\eqref{eq:two_interacting_harmonic_oscillators} describing two coupled, driven harmonic oscillators was expressed in terms of the position and momenta operators corresponding to the two oscillators.
For numerical simulations, we will find it convenient to work with creation and annihilation operators instead, with the system and environment annihilation operators denoted by $\opm a$ and $\opm b$ respectively.
The annihilation ladder operators for the two oscillators can be expressed as~\cite{sakurai2020modern}
\bseq
\begin{align}
    \opm a (t) &= \sqrt{\frac{m\omega_a(t)}{2\hbar}} \left( \opm x_a +\frac{i}{m\omega_a(t)} \opm p_a\right), \label{eq:ladder_operator_a} \\
    \text{and }\opm b &= \sqrt{\frac{m\omega_b}{2\hbar}} \left( \opm x_b +\frac{i}{m\omega_b} \opm p_b\right).
\end{align}
\eseq
The operators $\opm a$ and $\opm a^\dagger$ are explicitly time-dependent due to the $\omega_a(t)$ terms.
Henceforth, we will drop the time dependence of $\opm a$ and $\omega_a$ for notational simplicity.

In terms of the ladder operators, the Hamiltonian takes the form
\bseq \label{eq:a_and_a_dagger_Hamiltonian_two_osci}
\begin{align}
    \htot  &= \opm{H}_S \otimes \eye + \opm{V}_{SE} + \eye \otimes \opm H_E \\
    \opm{H}_S&=\hbar \omega_a \left(\opm{a}^\dagger \opm{a}+\frac{1}{2} \right),\\
    \opm{H}_E&=\hbar \omega_b \left(\opm{b}^\dagger \opm{b}+\frac{1}{2} \right), \\
    \opm{V}_{SE}&=\frac{\hbar g_x }{m \sqrt{\omega_a \omega_b}}(\opm{a}+\opm{a}^\dagger )(\opm{b}+\opm{b}^\dagger). \label{eq:two_oscilators_VSE_ladders}
\end{align}
\eseq
[As expected, the only term above that is time dependent above is $\opm H_S$. 
Even though $\opm V_{SE}$ appears to be time-dependent, the explicit time-dependence of $\opm a$ and $\opm a^\dagger$ from Eq.~\eqref{eq:ladder_operator_a} cancels with the time dependence of the $\omega_a$ term in Eq.~\eqref{eq:two_oscilators_VSE_ladders}.]

The power and heat-flux operators [see Eq.~\eqref{eq:two_interacting_spins_wdot_qdot_xp}] can thus be equivalently expressed in terms of the ladder operators as
\bseq \label{eq:two_interacting_spins_wdot_qdot_ab}
\begin{align}
    \wdot &= \frac{\hbar \dot \omega_a}{2m} (\opm a ^\dagger + \opm a)^2,\\
    \text{and }\qdot &= -i\hbar \frac{g}{m} \sqrt{\frac{\omega_a}{\omega_b}} (\opm a^\dagger - \opm a) (\opm b ^\dagger + \opm b).
\end{align}
\eseq

We note that the Hamiltonian dynamics corresponding to  Eq.~\eqref{eq:a_and_a_dagger_Hamiltonian_two_osci} can be solved exactly using the techniques developed in Ref.~\cite{harmo2}. 
However, we restrict our analysis to a numerical study.
Similar to the previous example, we numerically simulate the evolution of two interacting oscillators and compare the obtained values of $\dv{\langle \opm U \rangle}{t}$ with those of $\langle \udot \rangle$ (see Fig.~\ref{subfig:udot_vs_dot_u_osci_sine}).
For this simulation, we considered a truncated Hilbert space consisting of the lowest $100$ energy levels of each of the two oscillators (so that the total Hilbert space dimension used was $100000$), and solved the Schr\"odinger equation with Hamiltonian Eq.~\eqref{eq:a_and_a_dagger_Hamiltonian_two_osci} truncated to these levels.
Consequently, while the two curves match, due to this approximation, we find that at later times, $\langle \udot \rangle$ deviates from the derivative of mean internal energy.
Another potential source of error could be the fact that the operator definitions $\wdot$, $\qdot$ and $\udot$ were based on the cyclic property of matrix traces, which does not always hold for infinite dimensional matrices.
However, for any finite but sufficiently large truncation of Eq.~\eqref{eq:a_and_a_dagger_Hamiltonian_two_osci}, we expect to Eq.~\eqref{eq:udot_and_dot_u} to be satisfied upto numerical errors.

\section{Bounds on \texorpdfstring{$\sigma_\udot$}{} in terms of \texorpdfstring{$\sigma_\wdot$}{} and \texorpdfstring{$\sigma_\qdot$}{}} \label{app:t_pm}
Here, we will obtain an upper bound on $\sigma_{\udot}^2$ in terms of $\sigma_{\wdot}$ and $\sigma_{\qdot}$.
First, we note that
\bseq
\begin{align}
    \udot &= \wdot + \qdot, \label{eq:Udot_is_sum}\\
    \text{and } \langle \udot \rangle &= \langle \wdot\rangle + \langle \qdot\rangle. \label{eq:Udot_first_moment}
\end{align}
\eseq
Taking the square of Eq.~\eqref{eq:Udot_is_sum} and subtracting from it the square of Eq.~\eqref{eq:Udot_first_moment}, we get
\begin{align}
    \langle \udot ^2 \rangle - \langle \udot \rangle^2 &= \langle \wdot^2 \rangle - \langle \wdot \rangle^2 + \langle \qdot^2 \rangle - \langle \qdot \rangle^2\\
    & \ \ \ \langle \{ \qdot, \wdot\}\rangle - 2 \langle \qdot \rangle \langle \wdot\rangle \nonumber \\
    \text{that is } \sigma_{\udot}^2 &= \sigma_{\wdot}^2 + \sigma_{\qdot}^2 + 2 \cov (\qdot, \wdot).
\end{align}
Adding and subtracting $2 \sigma_{\qdot} \sigma_{\wdot}$, we have
\begin{align}
    \sigma_\udot ^2 = (\sigma_\qdot + \sigma_\wdot) ^2 - 2 (\sigma_\qdot \sigma_\wdot - \cov (\qdot, \wdot)) \label{eq:sigma_square_udot_e1}
\end{align}
or alternatively,
\begin{align}
\sigma_\udot^2 &= (\sigma_\qdot - \sigma_\wdot)^2 + 2(\sigma_\qdot \sigma_\wdot + \cov (\qdot, \wdot)) \label{eq:sigma_square_udot_e2}
\end{align}
From this, we obtain
\begin{align}
    \sigma_\wdot \sigma_\qdot \pm \cov (\qdot , \wdot) &\geq  t_{\pm} \geq 0,
\end{align}
Eq.~\eqref{eq:sigma_square_udot_e1} and \eqref{eq:sigma_square_udot_e2} thus give us
\begin{align}
    (\sigma_\qdot - \sigma_\wdot)^2 + 2t_+ \leq \sigma_\udot ^2 \leq (\sigma_\qdot + \sigma_\wdot) ^2 - 2t_- \label{eq:bounds_sigma_udot}
\end{align}

\section{Exact expressions for the single qubit quantum battery example}\label{app:1qug}
In  this section, we obtain an exact expression for the commutator term for the energy-power uncertainty relation for the example of a single qubit quantum battery with the Hamiltonian Eq.~\eqref{eq:isolated_single_qubit_H}.
Let us first provide some further background on the expression in Eq.~\eqref{eq:energy_power_uncertainty_closed_battery} of the main text for generic quantum batteries.
If we write the matrices $\opm{V}_S(t)$ and $\opm{\rho}(t)$ in the basis of the battery Hamiltonian $\opm{H}_0$, retaining only the commutator term in Eq.~\eqref{eq:energy_power_uncertainty_closed_battery}, we have 
\bseq \label{eq:uncrt}
\begin{align}
\sigma_{\opm E_B} \sigma_{\opm P_B^c} & \geq \frac{|\sum_{i\neq j}(\epsilon_i-\epsilon_j)^2 V_{ij}\rho_{ji} | }{2\hbar }\\
&=\frac{ \abs{\sum_{i<j} (\epsilon_i-\epsilon_j)^2\Re(V_{ij}\rho_{ji})}}{\hbar},
\end{align}
\eseq
where $\epsilon_i$s denote the eigenvalues of $\opm H_0$, and $V_{ij}$ and $\rho_{ij}$ denote the matrix elements of $\opm V_S$(t) and $\opm \rho(t)$ respectively.
The uncertainty bound clearly depends on the amount of decoherence in the $\opm H_0$ basis.
We note that measurements in the $\opm{H}_0$ basis make the uncertainty of the energy $\sigma^2_{\opm E_B}=0$ just after the measurement.
However, since the battery power operator is not diagonal in this basis in general, we can still have $\sigma_{P_B}^2 \geq 0$, as seen in Fig.~\ref{fig:epbat1q}.
The right-hand side of Eq.~\eqref{eq:uncrt} is also seen to vanish as expected.

Assuming that at time $t=0$, the state of the qubit is given by $\opm \rho_0 = \frac{1}{2}(\eye + \vec \sigma. \vec \beta)$, using Eq.~\eqref{eq:uncrt}, we find that the commutator term in the uncertainty bound at time $t$ is given by 
\begin{align}
    \bsplit
   &  \frac{|\sum_{i\neq j}(\epsilon_i-\epsilon_j)^2 \rho_{ij} v_{ji}| }{2 } = 4 h_3 ^2  \Re{V_{12}\rho_{21}(t)} \\
    &= \frac{2h_3^2}{\alpha^2} \times \left[ h_3^2 (\beta_1 v_1+v_1-\beta_2 v_2)+(3 \beta_1+1) v_1^3 \right.  \\
    & + (3 \beta_1+1) v_1v_2^2+(\beta_1+1) v_1v_3^2 \\
    & +2 h_3 \left(\beta_3 \left(v_1^2+v_2^2\right)+v_3(\beta_1 v_1+v_1-\beta_2 v_2)\right)  \\
    & +v_1^2 (\beta_2 v_2+2 \beta_3 v_3)+v_2\left(\beta_2 v_2^2+2 \beta_3 v_2v_3-\beta_2 v_3^2\right)  \\
    &+ \left. \cos(2t \alpha) t_1 + \sin(2t\alpha) t_2 \right],
    \esplit \label{eq:formula_for_bound_isolated_qubit}
\end{align}
where 
\begin{align}
    \bsplit
    t_1 &= h_3^2 (3 \beta_1 v_1+v_1+\beta_2 v_2)+\\
    & (\beta_1+1) v_1^3+(\beta_1+1) v_1v_2^2+(3 \beta_1+1) v_1v_3^2 \\
    &-2 \beta_3 h_3 \left(v_1^2+v_2^2\right)+ 2 h_3 v_3(3 \beta_1 v_1+v_1+\beta_2 v_2) \\
    & -\left(v_1^2 (\beta_2 v_2+2 \beta_3 v_3)\right) \\
    &+v_2\left(-\beta_2 v_2^2-2 \beta_3 v_2v_3+\beta_2 v_3^2\right) \\
    \text{and }t_2 &= \alpha \left(2 \beta_3 v_1(h_3+v_3)+(2 \beta_1+1) \left(v_1^2+v_2^2\right)\right)
    \esplit
\end{align}

\section{Haar averages}
\subsection{Key formulae}\label{app:keyfhaar}
Let us introduce the key formulae for the Haar average used in the main text. First, we will use $\overline{X}$ to indicate the expectation value over the Unitary Haar measure of a certain quantity, where $\opm{U}$ is the unitary operator being averaged.
In particular, we use the following formulae \cite{Oliviero2021,caravelli2020random}:
\bseq
\begin{align}
    \overline{ \opm{1} }&=\opm{1}\\
    \hat {\mathcal R}^{1}(\opm{U})=\overline{\opm{U} \opm{A} \opm{U}^\dagger}&=\Tr(\opm{A}) \eye\\
    \overline{(\opm{U} \otimes \opm{U}) (\opm{A}\otimes \opm{B}) (\opm{U}^\dagger\otimes \opm{U}^\dagger)}&=\lambda_+(\opm{A}\otimes \opm{B}) \opm{\Pi}_++\lambda_- (\opm{A}\otimes \opm{B}) \opm{\Pi}_-
\end{align}
\eseq
where 
\begin{align}
    \opm{\Pi}_{\pm}=\frac{\eye^{\otimes 2}\pm \mathbb {S}}{2}
\end{align}
where $\mathbb {S}$ is the swap operator, and where
\begin{align}
    \lambda_\pm (\opm{A}\otimes \opm{B})=\frac{\text{Tr}((\opm{A}\otimes \opm{B})\opm{\Pi}_{\pm})}{\text{Tr}(\opm{\Pi}_{\pm})}.
\end{align}
which are the formulae we use in the following. In particular, we have $\hat {\mathcal R}^{2}(\opm{A})=\overline{(\opm{U} \otimes \opm{U}) (\opm{A}\otimes \opm{A}) (\opm{U}^\dagger\otimes \opm{U}^\dagger)}$
 
In the following, we will use the following identity, due to the cyclicity of the trace and the properties of a commutator
\begin{align}
    \Tr([\opm{A},[\opm{B},\opm{C}]]\opm{D})=\Tr([[\opm{D},\opm{A}],\opm{B}]]\opm{C})=\Tr([\opm{B},[\opm{A},\opm{D}]]\opm{C})\label{eq:cictr},
\end{align}
which can be derived in a few steps of cyclicity of the trace, and the last from flipping the commutators twice. 
This expression will be useful when averaging with twirls over the interaction $\opm{V}_S$, as we can write, if $\opm{A}=\opm{B}$, then
\begin{align}
    \Tr([\opm{H}_0,[\opm{H}_0,\opm{V}_*]]\opm{\rho})=\Tr([\opm{H}_0,[\opm{H}_0,\opm{\rho}]]\opm{V}_*)\label{eq:flipping}
\end{align}
which an the expression we use in the following. 

\subsection{Quantum battery}
Let us now write this expression term by term. First, we note that $|\Tr(\opm{A})|^2=\Tr(\opm{A})\Tr(\opm{A})^*=\Tr(\opm{A})\Tr(\opm{A}^\dagger)$, and $[\opm{A},\opm{B}]^\dagger=[\opm{B}^\dagger,\opm{A}^\dagger]$.
For the commutator term, we have for hermitian $\opm{A},\opm{B}$
\bseq
\begin{align}
    \abs{\frac{1}{2i}\langle [\opm{A},\opm{B}]\rangle}^2&=\frac{1}{4}\Tr(\opm{\rho}_t[\opm{A},\opm{B}])\Tr([\opm{A},\opm{B}]^\dagger\opm{\rho}_t^\dagger)\\
    &=\frac{1}{4}\Tr([\opm{A},\opm{B}]\opm{\rho}_t \otimes [\opm{B}^\dagger,\opm{A}^\dagger]\opm{\rho}_t)\\
    &=\frac{1}{4}\Tr\Big(([\opm{A},\opm{B}]\otimes  [\opm{B}^\dagger,\opm{A}^\dagger]) (\opm{\rho}_t \otimes \opm{\rho}_t )\Big)\\
    &=\frac{1}{4}\Tr\Big(([\opm{A},\opm{B}]\otimes  [\opm{B}^\dagger,\opm{A}^\dagger])_{-t} (\opm{\rho}_0 \otimes \opm{\rho}_0 )\Big)
\end{align}
\eseq
which is the expression we use in the following.

\subsubsection{Twirling the initial state}\label{app:closedav}
We assume a random initial state as a twirl of the form
\begin{align}
    \opm{\rho}_0^U=U \opm{\rho}_0 U^\dagger
\end{align}
It follows that
\begin{align}
    \overline{\abs{\frac{1}{2i}\langle [\opm{A},\opm{B}]\rangle}^2}=\frac{1}{4}\Tr\Big(([\opm{A},\opm{B}]\otimes  [\opm{B}^\dagger,\opm{A}^\dagger])_{-t} (\lambda_+(\opm{\rho_0}^{\otimes 2}) \opm{\Pi}_++\lambda_- (\opm{\rho_0}^{\otimes 2}) \opm{\Pi}_- )\Big)
\end{align}
We note that we have
\begin{align}
    \Tr(\opm{\Pi}_{\pm})=\frac{1}{2}\Big(\Tr(\eye)\pm \Tr(\mathbb{S})\Big)=\frac{d_S(d_S\pm 1)}{2}
\end{align}
and thus
\begin{align}
    \lambda_{\pm}(\opm{\rho_0}^{\otimes 2})=\frac{\Tr(\opm{\rho_0})^2\pm \Tr(\opm{\rho_0}^{2})}{d_S(d_S\pm 1)}=\frac{1\pm\mathcal {P}}{d_S(d_S\pm 1)}
\end{align}
where $\mathcal {P}$ is the purity of the initial state,
\begin{align}
    \mathcal P=\tr(\opm{\rho}_0^2).
\end{align}
We then have
\begin{align}
    \overline{\abs{\frac{1}{2i}\langle [\opm{A},\opm{B}]\rangle}^2}&=\frac{1}{4}\Tr\Big(([\opm{A},\opm{B}]\otimes  [\opm{B}^\dagger,\opm{A}^\dagger])_{-t}  \big((1+\mathcal P)\frac{\opm{\Pi}_+}{d_S(d_S+1)}+(1-\mathcal P) \frac{\opm{\Pi}_-}{d_S(d_S-1)} \big)\Big)
 \end{align}
 where $\ _{t}$ indicates that $U(t)[\opm{A},\opm{B}]U^\dagger(t)$ are time evolved individually.

 Using the fact that
 \begin{align}
     a \opm{\Pi}_++b\opm{\Pi}_-=\frac{a+b}{2} \eye+\frac{a-b}{2} \mathbb{S}=l_i\eye+l_s \mathbb{S}
 \end{align}
 then we can rewrite the expression above as
\begin{align}
   \overline{\abs{\frac{1}{2i}\langle [\opm{A},\opm{B}]\rangle}^2}&= \frac{l_i}{4}\Tr([\opm{A},\opm{B}])\Tr([\opm{B}^\dagger,\opm{A}^\dagger])+\frac{l_s}{4}\Tr([\opm{A},\opm{B}][\opm{B}^\dagger,\opm{A}^\dagger]).
\end{align}
Note however that we can reduce the expression above simply to
\begin{align}
   \overline{\abs{\frac{1}{2i}\langle [\opm{A},\opm{B}]\rangle}^2}&=
   \frac{l_s}{4}\Tr([\opm{A},\opm{B}][\opm{B}^\dagger,\opm{A}^\dagger])
\end{align}
where we used the fact the trace of a commutator is zero,
and where $$l_s=\frac{d_S \mathcal P-1}{d_S(d_S^2-1)}.$$ It is interesting to note that if $\mathcal P=1/d_s$, then the right-hand side becomes zero.
In the limit $d_S\gg 1$, we can then use the approximation 
\begin{align}
    \big((1+\mathcal P)\frac{\opm{\Pi}_+}{d_S(d_S+1)}+(1-\mathcal P) \frac{\opm{\Pi}_-}{d_S(d_S-1)} \big)\approx_{d_S\gg 1} \frac{1}{d_S^2}\Big(\eye +\mathcal P\mathbb{S}\Big).
\end{align}
and obtain the formula
\begin{align}
    \overline{\abs{\frac{1}{2i}\langle [\opm{A},\opm{B}]\rangle}^2}=\frac{\mathcal P}{(2d_S  )^2}\Tr([\opm{A},\opm{B}][\opm{B}^\dagger,\opm{A}^\dagger])
\end{align}
which is the formula we will use in the following.

\subsubsection{Closed battery}
Let us now use the formula above in the case of the closed quantum system, e.g. the quantum battery. We replace $\opm{B}=-\frac{i}{\hbar}[\opm{H}_0,\opm{V}_S]$, and $\opm{A}=\opm{H}_0$.
We have, averaging over the initial conditions, and noticing that $[\opm{H}_0,[\opm{H}_0,\opm{V}_S]]^\dagger=[\opm{H}_0,[\opm{H}_0,\opm{V}_S]]$, we have
\begin{align}
    \overline{\mathcal B}_\rho=\overline{\abs{\frac{1}{2i \hbar }\langle [\opm{H}_0,[\opm{H}_0,\opm{V}_S]\rangle}^2}=\frac{l_s}{(2\hbar)^2}\Tr([\opm{H}_0,[\opm{H}_0,\opm{V}_S]]^2)
\end{align}
Using the properties of traces, and after straightforward calculation, it is easy to see that
\begin{align}
    \Tr([\opm{H}_0,[\opm{H}_0,\opm{V}_S]]^2)=\Tr\Big(\opm{H}_0^2\big(6 \opm{V}_S\opm{H}_0^2 \opm{V}_S-8 (\opm{H}_0 \opm{V}_S)^2+2\opm{H}_0^2\opm{V}_S^2\big) \Big)
\end{align}
which is the expression we use in the main text.

\subsubsection{Twirling the interaction}\label{sec:closedtwirlint}
We now consider the twirl of the interaction potential $\opm{V}_S$, similarly to what done in Ref.~\cite{caravelli2020random}. Using the expression for hermitian $\opm{H}_0,\opm{V}_S$,
\bseq
\begin{align}
    \mathcal B&=\frac{1}{4}\Tr\Big(\big([\opm{H}_0,[\opm{H}_0,\opm{V}_S]]\otimes  [[\opm{V}_S^\dagger ,\opm{H}_0^\dagger],\opm{H}_0^\dagger]\big) (\opm{\rho}_t \otimes \opm{\rho}_t )\Big)\\
    &=\frac{1}{4}\Tr\Big(\big([\opm{H}_0,[\opm{H}_0,\opm{\rho}_t]]\otimes  [\opm{H}_0,[\opm{H}_0,\opm{\rho}_t]]\big) (\opm{V}_S \otimes \opm{V}_S )\Big)
\end{align}
\eseq
in which we used expression (\ref{eq:flipping}). Since this is exactly the same expression as before, with $\opm{\rho}\leftrightarrow \opm{V}_S$, we have that averaging over the unitary channel $U$ with $U^\dagger \opm{V}_S U$, we have
\begin{align}
\overline{\mathcal B}_V=\frac{m_s}{(2\hbar)^2} \Tr([\opm{H}_0,[\opm{H}_0,\opm{\rho}_t]]^2)    
\end{align}
where 
\begin{align}
    m_s=\frac{d_S \mathcal V-1}{d_S(d_S^2-1)},
\end{align}
with $\mathcal V=\tr(\opm{V}^2_S)$, and also we can immediately write 
\begin{align}
    \Tr([\opm{H}_0,[\opm{H}_0,\opm{\rho}_t]]^2)=\Tr\Big(\opm{H}_0^2\big(6 \opm{\rho}_t\opm{H}_0^2 \opm{\rho}_t-8 (\opm{H}_0 \opm{\rho}_t)^2+2\opm{H}_0^2\opm{\rho}_t^2\big) \Big).
\end{align}
Thus, unlike the twirling over the initial state, we have that $\overline{\mathcal B}_V$ is still time dependent.

\subsection{Bath-System interactions}\label{sec:bathsystem}
Let us now consider the average of the right-hand side of the commutator, with a bit more structure. We assume that $\opm{B}=[\opm{B}_0,[\opm{B}_0,\opm{C_{ S}}\otimes \opm{C}_{E}]]$. We assume
now that $\opm{C}_{E}\rightarrow U\opm{C}_{E}U^\dagger$. 

Now let us assume that our operator $\opm{D}=\opm{D}_S\otimes \opm{D}_E$, $\opm{A}=\opm{B}=\opm{H}_0\otimes \opm{1}_E$, and $\opm{C}=\opm{C}_S\otimes \opm{C}_E$.
It is easy to see then that we can write, following eqn. (\ref{eq:cictr}):
\begin{align}
\Tr([[\opm{D},\opm{A}],\opm{B}]]\opm{C})=\Tr_{SE}([[\opm{D}_S,\opm{H}_0],\opm{H}_0]]\opm{C}_S \otimes\opm{D}_E\opm{C}_E)=\Tr_S([[\opm{D}_S,\opm{H}_0],\opm{H}_0]]\opm{C}_S )\Tr_E(\opm{D}_E\opm{C}_E)
\end{align}
In the equation above, we kept the notation to make clear that the traces are picked on different subspaces, but we will remove this extra notation in what follows.
We can see that the right hand side of the uncertainty relationship is the case of the closed quantum battery multiplied by a factor associated with the environment, which we call $\mathcal D_E$.

We choose $\opm{D}_S=\opm{\rho}_{S}(t)$, while $\opm{D}_E=\opm{\rho}_{E}$ is thermal, and we assume time-independent. Similarly, the interaction is given by $\opm{C}=\opm{V}_S\otimes \opm{V}_E$.

\subsubsection{Twirling the system initial state}\label{eq:twirlingintst}
Let us first note that twirling the initial density matrix in this setup is similar to the analysis performed in the previous section. We have
\bseq
\begin{align}
\overline{\mathcal B}_\rho=\overline{\abs{\frac{1}{2i}\langle [\opm{A},\opm{B}]\rangle}^2}&=\frac{1}{4}\overline{\Tr\Big([\opm{H_0},[\opm{H_0},\opm{V}_S]]U_t^\dagger U^\dagger\opm{\rho}_{0S} U U_t\Big)^2}\Tr\Big(\opm{\rho}_E  \opm{V}_E \Big)^2\\
&=\frac{l_s}{(2 \hbar )^2}\Tr([\opm{H}_0,[\opm{H}_0,\opm{V}_S]]^2)\Tr\Big(\opm{\rho}_E  \opm{V}_E \Big)^2
\end{align}
\eseq
which is similar to the expression we had before, but now a factor due to the average of the interaction with the bath enters. We note that $\Tr([\opm{H}_0,[\opm{H}_0,\opm{V}_S]]^2)=\|[\opm{H}_0,[\opm{H}_0,\opm{V}_S]]\|_F^2$.

\subsubsection{Twirling the interacting potential}
We now assume that the unitary channel operates as $\opm{V}_E\rightarrow U \opm{V}_E U^\dagger$.
\begin{align}
\overline{\mathcal B}_V=\overline{\abs{\frac{1}{2i}\langle [\opm{A},\opm{B}]\rangle}^2}&=\frac{1}{4}|\Tr\Big([\opm{H_0},[\opm{H_0},\opm{V}_S]]\opm{\rho}_{S}(t)\Big)|^2\cdot\overline{|\Tr\Big(\opm{\rho}_E U \opm{V}_E U^\dagger\Big)|^2}\\
&=\frac{\overline{\mathcal M_E}}{4}|\Tr\Big([\opm{H_0},[\opm{H_0},\opm{V}_S]]\opm{\rho}_{S}(t)\Big)|^2
\end{align}
We see from the equation above that the uncertainty is the same as the one of a quantum battery, but with a factor that depends on the bath $\mathcal M_E$. It is interesting to note that, because of the invariance of the Haar measure, and/or because of the ciclity of the trace, this is the same as averaging via $\opm{\rho}_E\rightarrow U^\dagger \opm{\rho}_E U$, while keeping $\opm{V}_E$ constant. For the average of this term, we can use the formulae from the previous section, to obtain
\begin{align}
\overline{\mathcal M_E}=\overline{\Tr\Big(\opm{\rho}_E U \opm{V}_E U^\dagger\Big)^2}=\frac{\Tr(\opm{V}_S^{\otimes 2} \opm{\Pi}^+)(1+\mathcal P_E)}{d_E(d_E+1)}+\frac{\Tr(\opm{V}_S^{\otimes 2} \opm{\Pi}^+)(1-\mathcal P_E)}{d_E(d_E-1)}
\end{align}
where $\mathcal P_E=\frac{1}{Z^2}\Tr(e^{-2\beta \opm{H}_E})$ is the purity of the bath. For $\beta=0$, we have $\mathcal P_E=1/d_E$, while for $\beta=\infty$, $\mathcal P_E=1$.
Note that in the limit of a large bath, we obtain, using the fact that $\opm{\Pi}^++\opm{\Pi}^-=\eye_E$,  and $\opm{\Pi}^+-\opm{\Pi}^-=\mathbb{S}_E$,
\begin{align}
\overline{\mathcal M_E}=\overline{\Tr\Big(\opm{\rho}_E U \opm{V}_E U^\dagger\Big)^2}\approx_{d_E\gg 1}\frac{1}{d_E^2}\Big(
\Tr(\opm{V}_E^{\otimes2}) +\Tr(\opm{V}_E^2) \mathcal P_E\Big)&=&\frac{1}{d_E^2}\Big(
\Tr(\opm{V}_E)^2 +\Tr(\opm{V}_E^2) \mathcal P_E\Big)\\
\end{align}
We thus obtain that depending on the interaction potential, the uncertainty is renormalized by an interaction with the bath and the temperature.
If we define the variance of an operator in the infinite temperature bath, e.g.
\begin{align}
    \Delta V^2_{\infty}=\frac{1}{d_E}\Big(\Tr(\opm{V}_E^2)-\Tr(\opm{V}_E)^2\Big),
\end{align} 
then we have
\begin{align}
    \overline{\mathcal M_E}=\frac{\Tr(\opm{V}_E^2)(1+\mathcal P_E)-d_E \Delta V^2_\infty}{d_E^2}
\end{align}
where $\mathcal P_E$ is the purity of the bath.

\end{appendix}
%\bibliography{biblio}

\begin{thebibliography}{10}
\providecommand{\url}[1]{\texttt{#1}}
\providecommand{\urlprefix}{URL }
\expandafter\ifx\csname urlstyle\endcsname\relax
  \providecommand{\doi}[1]{doi:\discretionary{}{}{}#1}\else
  \providecommand{\doi}{doi:\discretionary{}{}{}\begingroup
  \urlstyle{rm}\Url}\fi
\providecommand{\eprint}[2][]{\url{#2}}

\bibitem{KosloffQuantum2013}
R.~Kosloff,
\newblock \emph{Quantum {{Thermodynamics}}: {{A Dynamical Viewpoint}}},
\newblock Entropy \textbf{15}(6), 2100 (2013),
\newblock \doi{10.3390/e15062100}.

\bibitem{VinjanampathyQuantum2016}
S.~Vinjanampathy and J.~Anders,
\newblock \emph{Quantum thermodynamics},
\newblock Contemporary Physics \textbf{57}(4), 545 (2016),
\newblock \doi{10.1080/00107514.2016.1201896}.

\bibitem{AlickiIntroduction2018a}
R.~Alicki and R.~Kosloff,
\newblock \emph{Introduction to {{Quantum Thermodynamics}}: {{History}} and
  {{Prospects}}},
\newblock In F.~Binder, L.~A. Correa, C.~Gogolin, J.~Anders and G.~Adesso,
  eds., \emph{Thermodynamics in the {{Quantum Regime}}: {{Fundamental Aspects}}
  and {{New Directions}}}, pp. 1--33. Springer International Publishing, Cham,
\newblock ISBN 978-3-319-99046-0,
\newblock \doi{10.1007/978-3-319-99046-0_1} (2018).

\bibitem{StrasbergQuantum2022}
P.~Strasberg,
\newblock \emph{Quantum {{Stochastic Thermodynamics}}: {{Foundations}} and
  {{Selected Applications}}},
\newblock Oxford {{Graduate Texts}}. Oxford University Press, Oxford, New York,
\newblock ISBN 978-0-19-289558-5 (2022).

\bibitem{BochkovGeneral1977}
G.~N. Bochkov and E.~Kuzovlev,
\newblock \emph{General theory of thermal fluctuations in nonlinear systems},
\newblock Zh. Eksp. Teor. Fiz \textbf{72}, 238 (1977).

\bibitem{JarzynskiNonequilibrium1997}
C.~Jarzynski,
\newblock \emph{Nonequilibrium {{Equality}} for {{Free Energy Differences}}},
\newblock Physical Review Letters \textbf{78}(14), 2690 (1997),
\newblock \doi{10.1103/PhysRevLett.78.2690},
\newblock \url{https://link.aps.org/doi/10.1103/PhysRevLett.78.2690}.

\bibitem{JarzynskiEquilibrium1997}
C.~Jarzynski,
\newblock \emph{Equilibrium free-energy differences from nonequilibrium
  measurements: {{A}} master-equation approach},
\newblock Physical Review E \textbf{56}(5), 5018 (1997),
\newblock \doi{10.1103/PhysRevE.56.5018},
\newblock \url{https://link.aps.org/doi/10.1103/PhysRevE.56.5018}.

\bibitem{CrooksEntropy1999}
G.~E. Crooks,
\newblock \emph{Entropy production fluctuation theorem and the nonequilibrium
  work relation for free energy differences},
\newblock Physical Review E \textbf{60}(3), 2721 (1999),
\newblock \doi{10.1103/PhysRevE.60.2721},
\newblock \url{https://link.aps.org/doi/10.1103/PhysRevE.60.2721}.

\bibitem{KurchanQuantum2001}
J.~Kurchan,
\newblock \emph{A {{Quantum Fluctuation Theorem}}},
\newblock \url{http://arxiv.org/abs/cond-mat/0007360},
\newblock \doi{10.48550/arXiv.cond-mat/0007360} (2001),
  \eprint{cond-mat/0007360}.

\bibitem{TasakiJarzynski2000}
H.~Tasaki,
\newblock \emph{Jarzynski {{Relations}} for {{Quantum Systems}} and {{Some
  Applications}}},
\newblock \url{http://arxiv.org/abs/cond-mat/0009244},
\newblock \doi{10.48550/arXiv.cond-mat/0009244} (2000),
  \eprint{cond-mat/0009244}.

\bibitem{CampisiQuantum2011}
M.~Campisi, P.~Talkner and P.~H{\"a}nggi,
\newblock \emph{Quantum {{Bochkov}}--{{Kuzovlev}} work fluctuation theorems},
\newblock Philosophical Transactions of the Royal Society A: Mathematical,
  Physical and Engineering Sciences \textbf{369}(1935), 291 (2011),
\newblock \doi{10.1098/rsta.2010.0252}.

\bibitem{EspositoNonequilibrium2009a}
M.~Esposito, U.~Harbola and S.~Mukamel,
\newblock \emph{Nonequilibrium fluctuations, fluctuation theorems, and counting
  statistics in quantum systems},
\newblock Reviews of Modern Physics \textbf{81}(4), 1665 (2009),
\newblock \doi{10.1103/RevModPhys.81.1665},
\newblock \url{https://link.aps.org/doi/10.1103/RevModPhys.81.1665}.

\bibitem{CampisiColloquium2011}
M.~Campisi, P.~H{\"a}nggi and P.~Talkner,
\newblock \emph{{\emph{Colloquium}} : {{Quantum}} fluctuation relations:
  {{Foundations}} and applications},
\newblock Reviews of Modern Physics \textbf{83}(3), 771 (2011),
\newblock \doi{10.1103/RevModPhys.83.771},
\newblock \url{https://link.aps.org/doi/10.1103/RevModPhys.83.771}.

\bibitem{callen1991thermodynamics}
H.~B. Callen,
\newblock \emph{Thermodynamics and an Introduction to Thermostatistics},
\newblock John wiley \& sons (1991).

\bibitem{TalknerFluctuation2007}
P.~Talkner, E.~Lutz and P.~H{\"a}nggi,
\newblock \emph{Fluctuation theorems: {{Work}} is not an observable},
\newblock Physical Review E \textbf{75}(5), 050102 (2007),
\newblock \doi{10.1103/PhysRevE.75.050102},
\newblock \url{https://link.aps.org/doi/10.1103/PhysRevE.75.050102}.

\bibitem{RoncagliaWork2014}
A.~J. Roncaglia, F.~Cerisola and J.~P. Paz,
\newblock \emph{Work {{Measurement}} as a {{Generalized Quantum Measurement}}},
\newblock Physical Review Letters \textbf{113}(25), 250601 (2014),
\newblock \doi{10.1103/PhysRevLett.113.250601},
\newblock \url{https://link.aps.org/doi/10.1103/PhysRevLett.113.250601}.

\bibitem{DeffnerQuantum2016}
S.~Deffner, J.~P. Paz and W.~H. Zurek,
\newblock \emph{Quantum work and the thermodynamic cost of quantum
  measurements},
\newblock Physical Review E \textbf{94}(1), 010103 (2016),
\newblock \doi{10.1103/PhysRevE.94.010103}.

\bibitem{Perarnau-LlobetNoGo2017}
M.~{Perarnau-Llobet}, E.~B{\"a}umer, K.~V. Hovhannisyan, M.~Huber and A.~Acin,
\newblock \emph{No-{{Go Theorem}} for the {{Characterization}} of {{Work
  Fluctuations}} in {{Coherent Quantum Systems}}},
\newblock Physical Review Letters \textbf{118}(7), 070601 (2017),
\newblock \doi{10.1103/PhysRevLett.118.070601},
\newblock \url{https://link.aps.org/doi/10.1103/PhysRevLett.118.070601}.

\bibitem{Mukamel2003}
S.~Mukamel,
\newblock \emph{Quantum extension of the jarzynski relation: Analogy with
  stochastic dephasing},
\newblock Physical Review Letters \textbf{90}(17) (2003),
\newblock \doi{10.1103/physrevlett.90.170604}.

\bibitem{Monnai2005}
T.~Monnai,
\newblock \emph{Unified treatment of the quantum fluctuation theorem and the
  jarzynski equality in terms of microscopic reversibility},
\newblock Physical Review E \textbf{72}(2) (2005),
\newblock \doi{10.1103/physreve.72.027102}.

\bibitem{Allahverdyan2005}
A.~E. Allahverdyan and T.~M. Nieuwenhuizen,
\newblock \emph{Fluctuations of work from quantum subensembles: The case
  against quantum work-fluctuation theorems},
\newblock Physical Review E \textbf{71}(6) (2005),
\newblock \doi{10.1103/physreve.71.066102}.

\bibitem{Alickiquantum1979a}
R.~Alicki,
\newblock \emph{The quantum open system as a model of the heat engine},
\newblock Journal of Physics A: Mathematical and General \textbf{12}(5), L103
  (1979),
\newblock \doi{10.1088/0305-4470/12/5/007}.

\bibitem{SpohnIrreversible1978}
H.~Spohn and J.~L. Lebowitz,
\newblock \emph{Irreversible {{Thermodynamics}} for {{Quantum Systems Weakly
  Coupled}} to {{Thermal Reservoirs}}},
\newblock In \emph{Advances in {{Chemical Physics}}}, pp. 109--142. John Wiley
  \& Sons, Ltd,
\newblock ISBN 978-0-470-14257-8,
\newblock \doi{10.1002/9780470142578.ch2},
\newblock
  \url{https://onlinelibrary.wiley.com/doi/abs/10.1002/9780470142578.ch2}
  (1978).

\bibitem{Koslofflinear1984}
R.~Kosloff and M.~A. Ratner,
\newblock \emph{Beyond linear response: {{Line}} shapes for coupled spins or
  oscillators via direct calculation of dissipated power},
\newblock The Journal of Chemical Physics \textbf{80}(6), 2352 (1984),
\newblock \doi{10.1063/1.446987},
\newblock
  \url{https://pubs.aip.org/jcp/article/80/6/2352/154275/Beyond-linear-response-Line-shapes-for-coupled}.

\bibitem{Kosloffquantum1984}
R.~Kosloff,
\newblock \emph{A quantum mechanical open system as a model of a heat engine},
\newblock The Journal of Chemical Physics \textbf{80}(4), 1625 (1984),
\newblock \doi{10.1063/1.446862},
\newblock \url{https://doi.org/10.1063/1.446862}.

\bibitem{VenkateshQuantum2015}
B.~P. Venkatesh, G.~Watanabe and P.~Talkner,
\newblock \emph{Quantum fluctuation theorems and power measurements},
\newblock New Journal of Physics \textbf{17}(7), 075018 (2015),
\newblock \doi{10.1088/1367-2630/17/7/075018},
\newblock \url{https://dx.doi.org/10.1088/1367-2630/17/7/075018}.

\bibitem{LiWork2013}
H.~Li and J.-S. Wang,
\newblock \emph{Work distribution under continuous quantum histories},
\newblock \url{http://arxiv.org/abs/1304.6286},
\newblock \doi{10.48550/arXiv.1304.6286} (2013), \eprint{1304.6286}.

\bibitem{SolinasWork2013}
P.~Solinas, D.~V. Averin and J.~P. Pekola,
\newblock \emph{Work and its fluctuations in a driven quantum system},
\newblock Physical Review B \textbf{87}(6), 060508 (2013),
\newblock \doi{10.1103/PhysRevB.87.060508},
\newblock \url{https://link.aps.org/doi/10.1103/PhysRevB.87.060508}.

\bibitem{BaratoThermodynamic2015a}
A.~C. Barato and U.~Seifert,
\newblock \emph{Thermodynamic {{Uncertainty Relation}} for {{Biomolecular
  Processes}}},
\newblock Physical Review Letters \textbf{114}(15), 158101 (2015),
\newblock \doi{10.1103/PhysRevLett.114.158101}.

\bibitem{GingrichDissipation2016}
T.~R. Gingrich, J.~M. Horowitz, N.~Perunov and J.~L. England,
\newblock \emph{Dissipation {{Bounds All Steady-State Current Fluctuations}}},
\newblock Physical Review Letters \textbf{116}(12), 120601 (2016),
\newblock \doi{10.1103/PhysRevLett.116.120601}.

\bibitem{PietzonkaFinitetime2017}
P.~Pietzonka, F.~Ritort and U.~Seifert,
\newblock \emph{Finite-time generalization of the thermodynamic uncertainty
  relation},
\newblock Physical Review E \textbf{96}(1), 012101 (2017),
\newblock \doi{10.1103/PhysRevE.96.012101}.

\bibitem{HorowitzThermodynamic2020}
J.~M. Horowitz and T.~R. Gingrich,
\newblock \emph{Thermodynamic uncertainty relations constrain non-equilibrium
  fluctuations},
\newblock Nature Physics \textbf{16}(1), 15 (2020),
\newblock \doi{10.1038/s41567-019-0702-6}.

\bibitem{CarolloUnraveling2019}
F.~Carollo, R.~L. Jack and J.~P. Garrahan,
\newblock \emph{Unraveling the {{Large Deviation Statistics}} of {{Markovian
  Open Quantum Systems}}},
\newblock Physical Review Letters \textbf{122}(13), 130605 (2019),
\newblock \doi{10.1103/PhysRevLett.122.130605},
\newblock \url{https://link.aps.org/doi/10.1103/PhysRevLett.122.130605}.

\bibitem{HasegawaQuantum2020}
Y.~Hasegawa,
\newblock \emph{Quantum {{Thermodynamic Uncertainty Relation}} for {{Continuous
  Measurement}}},
\newblock Physical Review Letters \textbf{125}(5), 050601 (2020),
\newblock \doi{10.1103/PhysRevLett.125.050601}.

\bibitem{HasegawaThermodynamic2021}
Y.~Hasegawa,
\newblock \emph{Thermodynamic {{Uncertainty Relation}} for {{General Open
  Quantum Systems}}},
\newblock Physical Review Letters \textbf{126}(1), 010602 (2021),
\newblock \doi{10.1103/PhysRevLett.126.010602},
\newblock \url{https://link.aps.org/doi/10.1103/PhysRevLett.126.010602}.

\bibitem{hasegawa2023unifying}
Y.~Hasegawa,
\newblock \emph{Unifying speed limit, thermodynamic uncertainty relation and
  {Heisenberg} principle via bulk-boundary correspondence},
\newblock Nat. Commun. \textbf{14}(1), 2828 (2023),
\newblock \doi{10.1038/s41467-023-38074-8}.

\bibitem{CampaioliColloquium2023}
F.~Campaioli, S.~Gherardini, J.~Q. Quach, M.~Polini and G.~M. Andolina,
\newblock \emph{Colloquium: {{Quantum Batteries}}} (2023), \eprint{2308.02277}.

\bibitem{AlickiEntanglement2013}
R.~Alicki and M.~Fannes,
\newblock \emph{Entanglement boost for extractable work from ensembles of
  quantum batteries},
\newblock Physical Review E \textbf{87}(4), 042123 (2013),
\newblock \doi{10.1103/PhysRevE.87.042123}.

\bibitem{BinderQuantacell2015}
F.~C. Binder, S.~Vinjanampathy, K.~Modi and J.~Goold,
\newblock \emph{Quantacell: Powerful charging of quantum batteries},
\newblock New Journal of Physics \textbf{17}(7), 075015 (2015),
\newblock \doi{10.1088/1367-2630/17/7/075015}.

\bibitem{CampaioliEnhancing2017}
F.~Campaioli, F.~A. Pollock, F.~C. Binder, L.~C{\'e}leri, J.~Goold,
  S.~Vinjanampathy and K.~Modi,
\newblock \emph{Enhancing the {{Charging Power}} of {{Quantum Batteries}}},
\newblock Physical Review Letters \textbf{118}(15), 150601 (2017),
\newblock \doi{10.1103/PhysRevLett.118.150601}.

\bibitem{AndolinaChargermediated2018}
G.~M. Andolina, D.~Farina, A.~Mari, V.~Pellegrini, V.~Giovannetti and
  M.~Polini,
\newblock \emph{Charger-mediated energy transfer in exactly solvable models for
  quantum batteries},
\newblock Physical Review B \textbf{98}(20), 205423 (2018),
\newblock \doi{10.1103/PhysRevB.98.205423}.

\bibitem{RossiniQuantum2020}
D.~Rossini, G.~M. Andolina, D.~Rosa, M.~Carrega and M.~Polini,
\newblock \emph{Quantum {{Advantage}} in the {{Charging Process}} of
  {{Sachdev-Ye-Kitaev Batteries}}},
\newblock Physical Review Letters \textbf{125}(23), 236402 (2020),
\newblock \doi{10.1103/PhysRevLett.125.236402}.

\bibitem{Julia-FarreBounds2020a}
S.~{Juli{\`a}-Farr{\'e}}, T.~Salamon, A.~Riera, M.~N. Bera and M.~Lewenstein,
\newblock \emph{Bounds on the capacity and power of quantum batteries},
\newblock Physical Review Research \textbf{2}(2), 023113 (2020),
\newblock \doi{10.1103/PhysRevResearch.2.023113}.

\bibitem{GyhmQuantum2022a}
J.-Y. Gyhm, D.~{\v S}afr{\'a}nek and D.~Rosa,
\newblock \emph{Quantum {{Charging Advantage Cannot Be Extensive}} without
  {{Global Operations}}},
\newblock Physical Review Letters \textbf{128}(14), 140501 (2022),
\newblock \doi{10.1103/PhysRevLett.128.140501},
\newblock \url{https://link.aps.org/doi/10.1103/PhysRevLett.128.140501}.

\bibitem{QuachSuperabsorption2022}
J.~Q. Quach, K.~E. McGhee, L.~Ganzer, D.~M. Rouse, B.~W. Lovett, E.~M. Gauger,
  J.~Keeling, G.~Cerullo, D.~G. Lidzey and T.~Virgili,
\newblock \emph{Superabsorption in an organic microcavity: {{Toward}} a quantum
  battery},
\newblock Science Advances \textbf{8}(2), eabk3160 (2022),
\newblock \doi{10.1126/sciadv.abk3160}.

\bibitem{JoshiExperimental2022}
J.~Joshi and T.~S. Mahesh,
\newblock \emph{Experimental investigation of a quantum battery using
  star-topology {{NMR}} spin systems},
\newblock Physical Review A \textbf{106}(4), 042601 (2022),
\newblock \doi{10.1103/PhysRevA.106.042601}.

\bibitem{RobertsonUncertainty1929}
H.~P. Robertson,
\newblock \emph{The {{Uncertainty Principle}}},
\newblock Physical Review \textbf{34}(1), 163 (1929),
\newblock \doi{10.1103/PhysRev.34.163},
\newblock \url{https://link.aps.org/doi/10.1103/PhysRev.34.163}.

\bibitem{schrodinger1930heisenbergschen}
E.~Schr{\"o}dinger,
\newblock \emph{Zum heisenbergschen unsch{\"a}rfeprinzip},
\newblock Akademie der Wissenschaften (1930).

\bibitem{Senuncertainty2014}
D.~Sen,
\newblock \emph{The uncertainty relations in quantum mechanics},
\newblock Current Science \textbf{107}(2), 203 (2014),
\newblock \url{https://www.jstor.org/stable/24103129},
\newblock \eprint{24103129}.

\bibitem{ScopaExact2019}
S.~Scopa, G.~T. Landi, A.~Hammoumi and D.~Karevski,
\newblock \emph{Exact solution of time-dependent {{Lindblad}} equations with
  closed algebras},
\newblock Physical Review A \textbf{99}(2), 022105 (2019),
\newblock \doi{10.1103/PhysRevA.99.022105}.

\bibitem{DannTimedependent2018}
R.~Dann, A.~Levy and R.~Kosloff,
\newblock \emph{Time-dependent {{Markovian}} quantum master equation},
\newblock Physical Review A \textbf{98}(5), 052129 (2018),
\newblock \doi{10.1103/PhysRevA.98.052129}.

\bibitem{LandiRevModPhys.93.035008}
G.~T. Landi and M.~Paternostro,
\newblock \emph{Irreversible entropy production: From classical to quantum},
\newblock Rev. Mod. Phys. \textbf{93}, 035008 (2021),
\newblock \doi{10.1103/RevModPhys.93.035008}.

\bibitem{JohanssonQuTiP2012}
J.~R. Johansson, P.~D. Nation and F.~Nori,
\newblock \emph{{{QuTiP}}: {{An}} open-source {{Python}} framework for the
  dynamics of open quantum systems},
\newblock Computer Physics Communications \textbf{183}(8), 1760 (2012),
\newblock \doi{10.1016/j.cpc.2012.02.021},
\newblock
  \url{https://www.sciencedirect.com/science/article/pii/S0010465512000835}.

\bibitem{JohanssonQuTiP2013}
J.~R. Johansson, P.~D. Nation and F.~Nori,
\newblock \emph{{{QuTiP}} 2: {{A Python}} framework for the dynamics of open
  quantum systems},
\newblock Computer Physics Communications \textbf{184}(4), 1234 (2013),
\newblock \doi{10.1016/j.cpc.2012.11.019},
\newblock
  \url{https://www.sciencedirect.com/science/article/pii/S0010465512003955}.

\bibitem{harmo0}
L.~E. Estes, T.~H. Keil and L.~M. Narducci,
\newblock \emph{Quantum-mechanical description of two coupled harmonic
  oscillators},
\newblock Physical Review \textbf{175}(1), 286–299 (1968),
\newblock \doi{10.1103/physrev.175.286}.

\bibitem{harmo1}
R.~Hab-arrih, A.~Jellal, D.~Stefanatos and E.~H. El~Kinani,
\newblock \emph{Virtual excitations and quantum correlations in ultra-strongly
  coupled harmonic oscillators under intrinsic decoherence},
\newblock Optik \textbf{278}, 170719 (2023),
\newblock \doi{10.1016/j.ijleo.2023.170719}.

\bibitem{harmo2}
D.~E. Bruschi, G.~S. Paraoanu, I.~Fuentes, F.~K. Wilhelm and A.~W. Schell,
\newblock \emph{General solution of the time evolution of two interacting
  harmonic oscillators},
\newblock Physical Review A \textbf{103}(2) (2021),
\newblock \doi{10.1103/physreva.103.023707}.

\bibitem{Nishiyama_2024}
T.~Nishiyama and Y.~Hasegawa,
\newblock \emph{Tradeoff relations in open quantum dynamics via robertson,
  maccone–pati, and robertson–schrödinger uncertainty relations},
\newblock Journal of Physics A: Mathematical and Theoretical \textbf{57}(41),
  415301 (2024),
\newblock \doi{10.1088/1751-8121/ad79cd}.

\bibitem{DongQuantum2022}
H.~Dong, D.~Reiche, J.-T. Hsiang and B.-L. Hu,
\newblock \emph{Quantum {{Thermodynamic Uncertainties}} in {{Nonequilibrium
  Systems}} from {{Robertson-Schr{\"o}dinger Relations}}},
\newblock Entropy \textbf{24}(7), 870 (2022),
\newblock \doi{10.3390/e24070870}.

\bibitem{Garcia-PintosFluctuations2020}
L.~P. {Garc{\'i}a-Pintos}, A.~Hamma and A.~{del Campo},
\newblock \emph{Fluctuations in {{Extractable Work Bound}} the {{Charging
  Power}} of {{Quantum Batteries}}},
\newblock Physical Review Letters \textbf{125}(4), 040601 (2020),
\newblock \doi{10.1103/PhysRevLett.125.040601},
\newblock \url{https://link.aps.org/doi/10.1103/PhysRevLett.125.040601}.

\bibitem{JarzynskiComparison2007}
C.~Jarzynski,
\newblock \emph{{Comparison of far-from-equilibrium work relations}},
\newblock Comptes Rendus. Physique \textbf{8}(5-6), 495 (2007),
\newblock \doi{10.1016/j.crhy.2007.04.010},
\newblock
  \url{https://comptes-rendus.academie-sciences.fr/physique/articles/10.1016/j.crhy.2007.04.010/}.

\bibitem{Baumgratz2014}
T.~Baumgratz, M.~Cramer and M.~Plenio,
\newblock \emph{Quantifying coherence},
\newblock Physical Review Letters \textbf{113}(14) (2014),
\newblock \doi{10.1103/physrevlett.113.140401}.

\bibitem{Caravelli2021}
F.~Caravelli, B.~Yan, L.~P. García-Pintos and A.~Hamma,
\newblock \emph{Energy storage and coherence in closed and open quantum
  batteries},
\newblock Quantum \textbf{5}, 505 (2021),
\newblock \doi{10.22331/q-2021-07-15-505}.

\bibitem{nielsen2001quantum}
M.~A. Nielsen and I.~L. Chuang,
\newblock \emph{Quantum computation and quantum information}, vol.~2,
\newblock Cambridge university press Cambridge (2001).

\bibitem{LeggettDynamics1987}
A.~J. Leggett, S.~Chakravarty, A.~T. Dorsey, M.~P.~A. Fisher, A.~Garg and
  W.~Zwerger,
\newblock \emph{Dynamics of the dissipative two-state system},
\newblock Reviews of Modern Physics \textbf{59}(1), 1 (1987),
\newblock \doi{10.1103/RevModPhys.59.1},
\newblock \url{https://link.aps.org/doi/10.1103/RevModPhys.59.1}.

\bibitem{schlosshauer2007quantum}
D.~Schlosshauer,
\newblock \emph{The quantum-to-classical transition},
\newblock The Frontiers Collection (Springer-Verlag, 2007)  (2007).

\bibitem{BlochNuclear1946}
F.~Bloch,
\newblock \emph{Nuclear {{Induction}}},
\newblock Physical Review \textbf{70}(7-8), 460 (1946),
\newblock \doi{10.1103/PhysRev.70.460},
\newblock \url{https://link.aps.org/doi/10.1103/PhysRev.70.460}.

\bibitem{TorreyTransient1949}
H.~C. Torrey,
\newblock \emph{Transient {{Nutations}} in {{Nuclear Magnetic Resonance}}},
\newblock Physical Review \textbf{76}(8), 1059 (1949),
\newblock \doi{10.1103/PhysRev.76.1059},
\newblock \url{https://link.aps.org/doi/10.1103/PhysRev.76.1059}.

\bibitem{SkinnerComprehensive2018}
T.~E. Skinner,
\newblock \emph{Comprehensive solutions to the {{Bloch}} equations and
  dynamical models for open two-level systems},
\newblock Physical Review A \textbf{97}(1), 013815 (2018),
\newblock \doi{10.1103/PhysRevA.97.013815},
\newblock \url{https://link.aps.org/doi/10.1103/PhysRevA.97.013815}.

\bibitem{brandao2012convergence}
F.~G. S.~L. Brand\~ao, P.~\ifmmode \acute{C}\else
  \'{C}\fi{}wikli\ifmmode~\acute{n}\else \'{n}\fi{}ski, M.~Horodecki,
  P.~Horodecki, J.~K. Korbicz and M.~Mozrzymas,
\newblock \emph{Convergence to equilibrium under a random {H}amiltonian},
\newblock Physical Review E \textbf{86}, 031101 (2012),
\newblock \doi{10.1103/PhysRevE.86.031101}.

\bibitem{caravelli2020random}
F.~Caravelli, G.~Coulter-De~Wit, L.~P. Garc\'{\i}a-Pintos and A.~Hamma,
\newblock \emph{Random quantum batteries},
\newblock Physical Review Research \textbf{2}, 023095 (2020),
\newblock \doi{10.1103/PhysRevResearch.2.023095}.

\bibitem{Caravelli2024}
F.~Caravelli,
\newblock \emph{Cycle equivalence classes, orthogonal weingarten calculus, and
  the mean field theory of memristive systems},
\newblock Neuromorphic Computing and Engineering \textbf{4}(2), 024005 (2024),
\newblock \doi{10.1088/2634-4386/ad4052}.

\bibitem{sakurai2020modern}
J.~J. Sakurai and J.~Napolitano,
\newblock \emph{Modern quantum mechanics},
\newblock Cambridge University Press (2020).

\bibitem{Oliviero2021}
S.~F.~E. Oliviero, L.~Leone, F.~Caravelli and A.~Hamma,
\newblock \emph{Random matrix theory of the isospectral twirling},
\newblock SciPost Physics \textbf{10}(3) (2021),
\newblock \doi{10.21468/scipostphys.10.3.076}.

\end{thebibliography}

\end{document}